\documentclass[usenatbib]{mn2e}
\usepackage{epsfig}
\usepackage{graphicx}
\usepackage{amssymb}
\usepackage{aas_macros}
\usepackage{color}

\newcommand{\kms}{\,{\rm km \, s^{-1}}}
\begin{document}

\title[Expected science yield from the {\it Gaia} Radial Velocity
Spectrometer]{Spectroscopic Survey of the Galaxy with {\it Gaia}
II. The expected science yield from the Radial Velocity Spectrometer}

\author[Wilkinson, Vallenari, Turon et
al.]{M.I.~Wilkinson$^{1}$\thanks{markw@ast.cam.ac.uk},
A.~Vallenari$^{2}$, C.~Turon$^{3}$, U.~Munari$^{4}$,D.~Katz$^{3}$
G.~Bono$^{5}$, \newauthor M.~Cropper$^{6}$, A.~Helmi$^{7}$,
N.~Robichon$^{3}$, F.~Th\'evenin$^{8}$, S.~Vidrih$^{9}$,
T.~Zwitter$^{9}$, \newauthor F.~Arenou$^{3}$, M.-O.~Baylac$^{3}$,
G.~Bertelli$^{2}$, A.~Bijaoui$^{8}$, F.~Boschi$^{4}$,
F.~Castelli$^{10,11}$, \newauthor F.~Crifo$^{3}$, M.~David$^{12}$,
A.~Gomboc$^{13,9}$, A.~G\'omez$^{3}$, M.~Haywood$^{3}$,
U.~Jauregi$^{9}$, \newauthor P.~de Laverny$^{8}$, Y.~Lebreton$^{3}$,
P.~Marrese$^{4}$, T.~Marsh$^{14}$, S.~Mignot$^{3}$, D.~Morin$^{3}$,
\newauthor S.~Pasetto$^{2}$, M.~Perryman$^{15}$, A.~Pr{\v s}a$^{9}$,
A.~Recio-Blanco$^{8}$, F.~Royer$^{16,3}$, A.~Sellier$^{3}$, \newauthor
A.~Siviero$^{4}$, R.~Sordo$^{4}$, C.~Soubiran$^{17}$,
L.~Tomasella$^{4}$, and Y.~Viala$^{3}$\\ $^1$Institute of Astronomy,
Madingley Road, Cambridge CB3 0HA, UK\\ $^2$INAF - Osservatorio
Astronomico di Padova, Vicolo Osservatorio 5, 35122 Padova, Italy\\
$^3$Observatoire de Paris, GEPI, 5 Place Jules Janssen, F-92195 Meudon
France\\ $^4$Osservatorio Astronomico di Padova INAF, sede di Asiago,
36012 Asiago (VI), Italy\\ $^5$INAF - Osservatorio Astronomico di
Roma, Via Frascati 33, 00040 Monte Porzio Catone, Italy\\ $^6$Mullard
Space Science Laboratory, University College London, Holmbury St Mary,
Dorking, Surrey RH5 6NT, UK\\ $^7$Kapteyn Astronomical Institute,
P.O.Box 800, 9700 AV Groningen, The Netherlands\\ $^8$O.C.A., BP 4229,
F-06304 Nice Cedex 4, France\\ $^9$University of Ljubljana, Dept. of
Physics, Jadranska 19, 1000 Ljubljana, Slovenia\\ $^{10}$CNR-Istituto
di Astrofisica Spaziale e Fisica Cosmica, Via del Fosso del Cavaliere,
00133, Roma, Italy\\ $^{11}$INAF-Osservatorio Astronomico di Trieste,
via G.B. Tiepolo 11, 34131 Trieste, Italy\\ $^{12}$University of
Antwerp, Middelheimlaan 1, B-2020 Antwerpen, Belgium\\
$^{13}$Astrophysics Research Institute, Liverpool John Moores
University, Twelve Quays House, Egerton Wharf, Birkenhead, CH41 1LD,
UK\\ $^{14}$Department of Physics, University of Warwick, Coventry CV4
7AL, UK\\ $^{15}$Research and Scientific Support Department of ESA,
ESTEC, Postbus 299, Keplerlaan 1, Noordwijk NL-2200 AG, The
Netherlands\\ $^{16}$Observatoire de Gen\`eve, 51 chemin des
Maillettes, CH-1290 Sauverny\\ $^{17}$Observatoire Aquitain des
Sciences de l'Univers, UMR 5804, 2 rue de l'Observatoire, 33270
Floirac, France\\}

\date{Received ; accepted }

 \maketitle  

\begin{abstract}
The {\it Gaia} mission is designed as a Galaxy explorer, and will
measure simultaneously, in a survey mode, the five or six phase space
parameters of all stars brighter than 20th magnitude, as well as
providing a description of their astrophysical characteristics.  These
measurements are obtained by combining an astrometric instrument with
micro-arcsecond capabilities, a photometric system giving the
magnitudes and colours in 15 bands and a medium resolution
spectrograph named the Radial Velocity Spectrometer (RVS).  The latter
instrument will produce spectra in the 848 to 874 nm wavelength range,
with a resolving power R = 11\,500, from which radial velocities,
rotational velocities, atmospheric parameters and abundances can be
derived. A companion paper~\citep{paperI} presents the characteristics
of the RVS and its performance. This paper details the outstanding
scientific impact of this important part of the {\it Gaia} satellite
on some key open questions in present day astrophysics. The unbiased
and simultaneous acquisition of multi-epoch radial velocities and
individual abundances of key elements in parallel with the astrometric
parameters is essential for the determination of the dynamical state
and formation history of our Galaxy. Moreover, for stars brighter than
V $\simeq$ 15, the resolving power of the RVS will give information
about most of the effects which influence the position of a star in
the Hertzsprung-Russell diagram, placing unprecedented constraints on
the age, internal structure and evolution of stars of all
types. Finally, the RVS multi-epoch observations are ideally suited to
the identification, classification and characterisation of the many
types of double, multiple and variable stars.
\end{abstract}

\begin{keywords}
binaries: spectroscopy -- stars: general -- Galaxy: evolution --
Galaxy: formation -- Galaxy: kinematics and dynamics -- Galaxy:
structure
\end{keywords}

%

\section{Introduction}
The {\it Gaia} mission was approved as one of the next cornerstones of
the European Space Agency (ESA) scientific programme in September
2000, and was confirmed as one of the flagship missions of ESA Cosmic
Vision in November 2003. The mission was conceived~\citep{LinPerr} in
the light of the success of {\it Hipparcos}, the first space
astrometry mission, also operated by ESA. This first mission proved
the very high potential of such an instrument by providing, for the
first time, a large data set of high-accuracy, homogeneous absolute
trigonometric parallaxes and proper motions. Milli-arcsecond level
accuracy was obtained for 118\,000 pre-selected stars. More than 3000
papers using this unique set of data have been published to date, in
domains ranging from reference frames to the cosmic distance scale,
through stellar physics and Galactic astronomy.

Three aspects of the {\it Gaia} mission represent major improvements
with respect to {\it Hipparcos}: an unprecedented astrometric accuracy
of a few micro-arcseconds, the systematic, and hence unbiased,
observation of some one billion objects down to a $G$
magnitude\footnote{The characteristics of the $G$ filter are discussed
in detail in ~\citeauthor{paperI} (\citeyear{paperI}; hereinafter
Paper~I). For example, a G5V star has $G-V = -0.36$.} of $20$, and the
simultaneous acquisition of astrometric parameters and of multi-epoch
multi-colour photometry and spectroscopic observations.  The overall
science case for {\it Gaia} is impressive and
multi-faceted~\citep[see][ and references therein]{Perr2001,bt02}. It
will clarify the origin and formation history of our Galaxy through
the provision of a quantitative stellar census, a kinematical and a
dynamical description of each of the Galactic stellar populations, the
identification of accretion debris, a determination of the Galactic
rotation curve and the distribution of dark matter. It will also
detect and characterize all types of stars, even the rarest types or
the more rapid evolutionary phases, placing unprecedented constraints
on the internal physics of stars, stellar evolution and stellar
ages. The {\it Gaia} results will also firmly establish the distance
scales in our Galaxy and far beyond and will detect tens of thousands
of double and multiple systems, variable stars, minor bodies in the
Solar System, quasars and supernovae. Thousands of extra-solar planets
will be systematically identified and their precise orbits and masses
determined. Finally, {\it Gaia} will perform a strong test of General
Relativity by measuring the Parametrized Post-Newtonian (PPN)
parameters $\gamma$ and $\beta$ with unrivalled precision.

The {\it Gaia} spectroscopic instrument, the Radial Velocity
Spectrometer (RVS), will contribute significantly to this science
case~\citep[see][ and references therein]{2003gsst.conf.....M}, by
providing spectra for 100 to 250 million stars up to magnitude V =
17. Its key contributions will be the following:
\begin{itemize}
\item the simultaneous acquisition of the third component of the space
velocity, the radial velocity, essential for kinematic and dynamical
studies in the Galaxy. These data will give direct insight into the
past accretion events which led to the formation of the Galaxy as it
is now observed, and will place strong observational constraints on
formation scenarios for the various components of the Galaxy;

\item multi-epoch measurements of radial velocities, essential for the
detection, classification and complete characterisation of binary
systems, and key observables for the determination of the fundamental
parameters of pulsating variable stars;

\item the determination of the `perspective acceleration', mandatory
for obtaining the expected astrometric accuracy for high velocity
stars;

\item the astrophysical characterisation of the brightest subset of
{\it Gaia} stars, complementary to the information provided by {\it
Gaia} photometric observations. In particular, the overall metal
abundance will be derived up to V $\approx$ 14-15 and individual
element abundances will be obtained up to V $\approx$ 12-13. For
fainter stars, photometry and astrometry will be used to estimate
abundance parameters -- RVS spectra will calibrate these estimates at
the bright end of the luminosity function.  These data will help to
resolve the age-metallicity degeneracy which currently hinders
interpretation of the colour-magnitude diagrams that are a fundamental
input to our understanding of the chemical evolution of the Galaxy;

\item the observation of the 862 nm Diffuse Interstellar Band which
will contribute to the construction of a three-dimensional map of
Galactic interstellar reddening.
\end{itemize}

The inclusion of the RVS on board the {\it Gaia} satellite has
critical advantages over any ground-based survey because it will
obtain spectra simultaneously with the astrometric parameters and
photometric indices, and for systematically the same stars. Even
though much effort was put into the planning of spectroscopic
observations of a large proportion of the {\it Hipparcos} target list
~\citep{gerbaldi89,mayor89} with particular emphasis on the
kinematically unbiased selection of the stars to be observed, the
large number of required observations has meant that the resulting
measurement programme is still incomplete and only parts of it have
been
published~\citep{grenier1999a,grenier1999b,2004A&A...418..989N}. The
absence of an un-biased radial velocity data set to complement the
{\it Hipparcos} astrometric observations constituted a severe
constraint on the value of the data set for the study of Galactic
kinematics. In fact, as~\cite{bdhmp97} emphasised, the non-uniformity
of published velocities for stars in the {\it Hipparcos} sample meant
that (in 1997) the set of stars with full three-dimensional velocity
information was kinematically biased because high proper motion
objects were more likely to have measured radial velocities. This had
a significant impact on the potential scientific value of this data
set. In the case of {\it Gaia}, the measurement principles ensure the
total homogeneity of the observations, with the sole exception of the
approximately $10$ percent of the densest regions of the Galaxy (the
cores of globular clusters and the inner Galactic bulge). After the
detection of particularly interesting objects, it will clearly be
fruitful to observe them with ground-based instruments providing much
higher spectral resolution.

\begin{table}
\caption{Precision ($1\sigma$ error) of the RVS instrument in the
determination of stellar parameters. The table gives the expected
accuracies for effective temperature $T_{\rm eff}$, surface gravity
$\log g$, iron abundance [Fe/H], and rotational velocity $v\sin
i$. For all parameters other than rotational velocity, the quoted
precisions will be obtained for a G5V star to a limiting magnitude of
V$\simeq 14.0$. Limiting magnitudes for other stellar types, as well
as a complete discussion of the algorithms used to calculate these
accuracy estimates, will be presented elsewhere (Recio-Blanco et al.,
in preparation). Paper~I also contains some additional discussion of
these performances.}
\label{tab:performance}
\begin{center}
\begin{tabular}{lrrrrrrr}
\hline
\multicolumn{8}{c}{Expected stellar parameter accuracies}\\\hline
\multicolumn{8}{l}{$\sigma(T_{\rm eff}) = 75 \pm 55$ K} \\
\multicolumn{8}{l}{$\sigma(\log g) = 0.16 \pm 0.12$ dex}\\
\multicolumn{8}{l}{$\sigma(\mbox{[Fe/H]}) = 0.11 \pm 0.08$ dex} \\
\multicolumn{8}{l}{$\sigma(v\sin i) = 5\kms$ to V=15 for late type stars}\\
\multicolumn{8}{l}{$\sigma(v\sin i) = 10-20\kms$ to V=10-11 for B5 main-sequence stars} \\\hline\hline
\end{tabular}
\end{center}
\end{table}

\begin{table}
\caption{Precision ($1\sigma$ error) of the RVS instrument in the
determination of radial velocities. The quantity $\sigma_1$ is the
$1\sigma$ error in radial velocity at the limiting magnitude $V_1$
after a single transit of the RVS instrument; $\sigma_2$ is the
$1\sigma$ error in radial velocity at the limiting magnitude $V_2$
after 102 transits (= average number of transits per star during the
mission); $V_3$ is the limiting magnitude for velocity errors of
$1\kms$ after 102 transits. Results for four stellar types are shown:
an F2 II, a G5V and a K1 III star each of solar metallicity and a
low-metallicity K1 III star. In each case, the limiting magnitude
$V_2$ is taken to be the faintest magnitude (to the nearest 0.5
magnitudes) at which the radial velocity error is $\lesssim20\kms$.
For more details see Paper~I.}
\label{tab:vel_performance}
\begin{center}
\begin{tabular}{lrrrrrrrr}
\hline
\multicolumn{8}{c}{Expected radial velocity accuracies}\\\hline

Type & [Fe/H] & $v\sin i$ & $\sigma_1$ & $V_1$ & $\sigma_2$ & $V_2$ & $V_3$ \\\hline
F2 II &	0.0 & 20 & 14.4 & 13.5 & 20.8 & 16.5 & 13.0\\
G5V & 0.0 & 5 & 19.7 & 14.5 & 16.0 & 17.0 & 14.0\\
K1 III & 0.0 & 5 & 11.5 & 15.0 & 10.6 & 17.5 & 14.0\\
K1 III & -1.5 & 5 & 11.7 & 14.5 & 10.3 & 17.0 & 14.0\\\hline
\end{tabular}
\end{center}
\end{table}

This paper summarises the expected impact of the RVS on our knowledge
of Galactic structure and evolution (\S~\ref{sec:gal_structure});
binary star statistics and characterisation (\S~\ref{sec:binaries});
stellar physics and evolution, as well as variable stars
(\S~\ref{sec:stellar_evolution}). In \S~\ref{sec:perform} we briefly
summarise the performance of the RVS in estimating velocities as a
function of position on the sky as this determines the contribution
which the RVS will make to our understanding of the various components
of the Galaxy. Tables~\ref{tab:performance} and~\ref{tab:vel_performance}
summarise the expected precisions and limiting magnitudes for various
stellar parameters including radial velocity. For more details of the
performance of the RVS instrument, the reader is directed to
Paper~I. In the present paper, we emphasise the most important
potential scientific contributions of the RVS. However, we note that
there will be other topics, in addition to those discussed here, for
which the RVS data set will be valuable.

\section{Galactic Structure and evolution}
\label{sec:gal_structure}

One of the key science drivers for the {\it Gaia} mission is to determine
the current dynamical state and formation history of the Milky
Way. Radial velocities provide the 6th phase space coordinate, vital
for building a complete picture of the stellar dynamics in the
Galaxy. In this regard, an unbiased sample of radial velocities is an
essential complement to the {\it Gaia} proper motion and distance data
set. The presence of the RVS on board the {\it Gaia} satellite will ensure
that multi-epoch radial velocities will be acquired in an un-biased
manner for all stars brighter than $V \sim 17$.

The most popular scenario for the formation of the Milky Way involves
the merger of several similar sized systems (whose exact number may
vary between 5 and 20) up to a redshift $z \sim 2 -
3$~\citep{kauffmann93,anse03}. These systems contribute both dark
matter, stars and gas. The accretion of smaller systems occurs
throughout the formation history of the Galaxy, but it must have been
the dominant form of mass growth since $z \sim 1$, when the thin disc
probably began to form. It is unclear whether the bulge of the Galaxy
was formed from the mergers of equal size objects at very high z, or
from the instability of a pre-existing disc. At high z, when mergers
were more frequent, it seems quite likely that discs could easily
become unstable, leading to the formation of a bar/bulge.  The mass
accreted at later times builds up the external halo and the thin disc
\citep{GilmoreWyse2001}. In recent years, considerable evidence of
past accretion events has been obtained (see
\S~\ref{sec:halo}). However, numerical simulations suggest that this
scenario for the formation of the Galaxy is not without its
problems. For example, there is an apparent discrepancy between the
hundreds of smaller sub-haloes surrounding a simulated Milky Way-type
galaxy in a standard $\Lambda$CDM universe and the $11$ dwarf
satellite galaxies which are observed to orbit the Milky
Way~\citep[e.g.][]{moore99}. \cite{anse03} also raise the concern that
simulated $\Lambda$CDM disc galaxies typically have significant bulge
components which may be difficult to reconcile with the observed
numbers of nearby, pure-disc galaxies. It is therefore important to
place observational constraints on the possible formation histories of
the various Galactic components. The most direct way to do this is to
look for correlations between the ages, metallicities and space
motions of their associated stellar populations.

In this section, we highlight some of the key Galactic stellar
populations which will be amenable to study with the RVS and the
scientific issues which can be addressed using the RVS data set. We
note that this is not intended to be an exhaustive review of Galactic
structure or of the contribution of the {\it Gaia} mission as a whole
to our understanding of the Galaxy, but rather a discussion of the key
contributions in this area to be expected from the RVS instrument.

\subsection{Expected RVS velocity performance}
\label{sec:perform}
\begin{figure*}
\resizebox{\hsize}{!}{\includegraphics{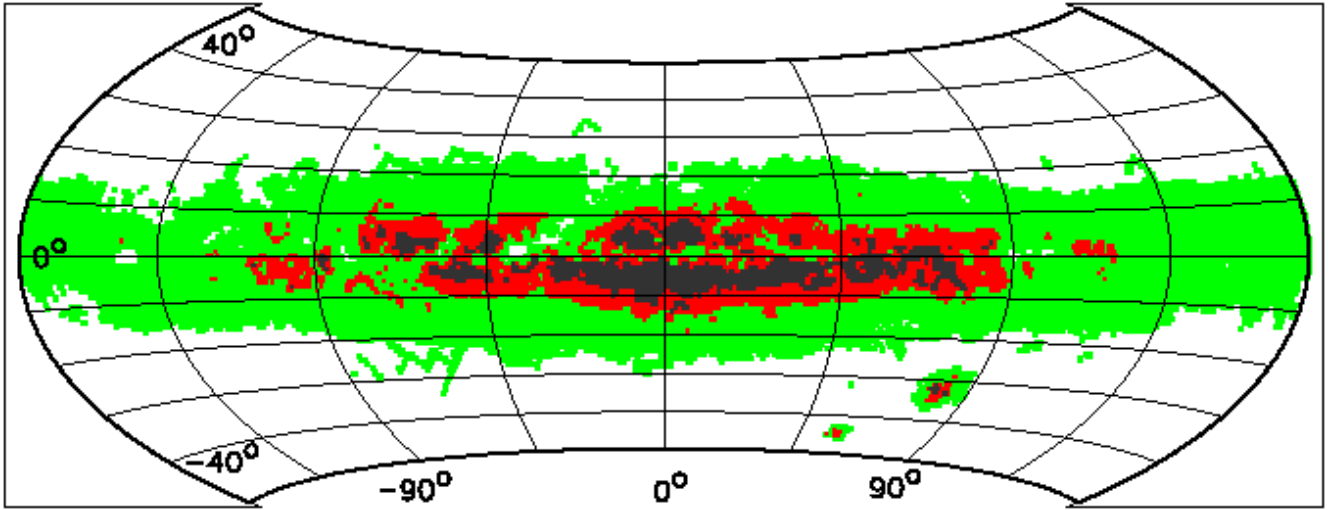}}
\caption{Stellar density as a function of Galactic coordinates ($l,b$)
for stars brighter than F$=17.5$ in the GSCII catalogue. Note that the
typical V$-$F colour of the disc stellar populations is about
0.6. Different colours correspond to regions in which the stellar
density is (i) less than $5\times 10^3$ deg$^{-2}$ (white) (ii)
$5\times 10^3 - 2\times 10^4$ deg$^{-2}$ (green) (iii) $2-4\times
10^4$ deg$^{-2}$ (red) and (iv) above $4\times 10^4$ deg$^{-2}$
(black). Current simulations show that the blue and green regions will
be fully observable by the RVS. However, the black regions will be
inaccessible as well as some of the red regions.}
\label{fig:l_b_density}
\end{figure*}

\begin{figure*}
\resizebox{\hsize}{!}{\includegraphics{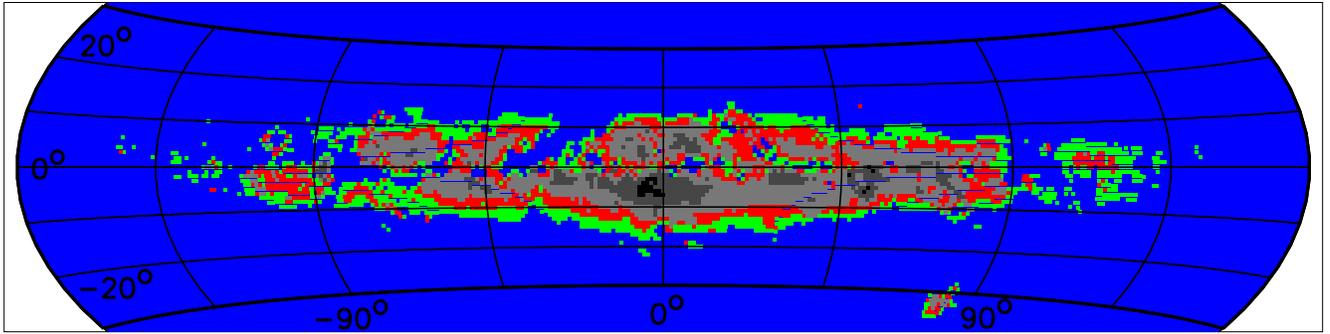}}
\caption{Expected limiting magnitude F$_{\rm lim}$ for RVS radial
velocities as a function of Galactic coordinates ($l,b$). The colours
correspond to limiting magnitudes of F$_{\rm lim}=17.5$ (blue),
F$_{\rm lim}=17.25$ (green), F$_{\rm lim}=16.75$ (red), F$_{\rm
lim}=16.25$ (light grey), F$_{\rm lim}=15.25$ (dark grey) and F$_{\rm
lim}=14.25$ (black). Note that this plot does not take into account
telemetry issues which will affect the densest regions of the
Galaxy. See text for discussion.}
\label{fig:l_b_mag_lim}
\end{figure*}

Fig.~\ref{fig:l_b_density} shows the stellar number density for all
stars brighter than F$=17.5$ as a function of Galactic coordinates $l$
and $b$ taken from the Guide Star Catalogue
II~\citep[GSCII:][]{gscii01}. This spectral band is close to that of
the RVS and in what follows we assume that these number densities are
representative of those which the RVS will encounter. As is discussed
in Paper~I, crowding at low Galactic latitudes severely limits the
ability of the RVS to provide robust velocity
measurements. Simulations of the effects of crowding
~\citep{2003gsst.conf..493Z} have demonstrated that degradation of
radial velocity determinations sets in when the number density exceeds
about $2\times 10^4$ stars per square degree. In the Figure, red and
black correspond to stellar densities above $2\times 10^4$ stars per
square degree and $4\times 10^4$ stars per square degree,
respectively. Thus, it is likely that the region of the sky within 10
degrees of the Galactic centre will not be observable with the
RVS. Techniques for the extraction of radial velocities from the data
in crowded fields are still being investigated, with the prospects of
some (limited) improvements over what is currently achievable. We also
note that the stellar densities in the Figure are averaged over tiles
of one square degree. It is well known~\citep[e.g.][]{popowski03} that
extinction levels towards the Galactic centre vary on small scales,
and so regions of high stellar density may be adjacent to regions of
significantly lower density. However, since the lowest extinction
regions (e.g. Baade's window) have stellar densities exceeding $10^6$
stars per square degree, Fig.~\ref{fig:l_b_density} probably gives a
realistic reflection of the regions in which the RVS will be able to
observe.

For stars of a given apparent magnitude F$_{\rm lim}$, simulations
have shown that, for late type stars, severe degradation of the radial
velocity accuracy occurs if the total stellar density of stars to a
limit roughly $0.75$ magnitude fainter than F$_{\rm lim}$ exceeds
$2\times 10^4$ stars per square degree. Thus, for a given region of
the sky, it is possible to determine an effective magnitude limit
above which we can expect the RVS to provide accurate radial
velocities. Fig.~\ref{fig:l_b_mag_lim} shows the effective limiting
magnitude for radial velocity measurements as a function of position
on the sky. This shows that over most of the sky the RVS will be able
to obtain velocities to a magnitude limit of between F$_{\rm lim}=16$
and F$_{\rm lim}=17$, although in the direction of the Galactic centre
this limit is significantly brighter. Given that the disc populations
have a mean ${\rm V-F}$ colour of 0.6, the limiting V magnitude will
range between $16.85$ to $17.5$ over most of the sky. In fact,
telemetric constraints will severely restrict the amount of data from
the Galactic centre region which can be obtained and so the magnitude
limit for these regions in Fig.~\ref{fig:l_b_mag_lim} is probably
optimistic. In the following, we assume that the RVS performances
described in Paper~I will be achieved over approximately $90$ per cent
of the sky, as suggested by Figs.~\ref{fig:l_b_mag_lim}
and~\ref{fig:l_b_density}.

\subsection{The Stellar Halo and Debris Streams}
\label{sec:halo} 

If the Milky Way was built via accretion as in the hierarchical
formation model~\citep[e.g.][]{wr78}, accretion events should have
left fossil signatures in its present day components, which should be
clearly detectable with {\it Gaia}~\citep{hw99,anse03}. The most
natural place to look for such substructures is the Galactic stellar
halo, since a spheroidal component is formed by the trails of stars
left by disrupted satellite
galaxies~\citep{Johnston95,morrison00}. Moreover, it is here where the
most metal-poor stars are found, which could imply that this component
was in place very early on~\citep{freeman02}.

Recent observations have shown that indeed considerable structure is
still present in the Milky Way's halo, indicating that accretion
events have had some role in its formation
history~\citep[e.g.][]{ibata94,majmunhaw96,helmi99,ivezic00}. The
existence of a substantial counter-rotating halo component is
suggested by RR-Lyrae and blue horizontal branch (BHB) star
kinematics~\citep{1992ApJS...78...87M,1999gaha.conf..230C}.  In the
Solar neighbourhood, it is unclear whether the local halo velocity
distribution is well approximated by a smooth
Gaussian~\citep{1998AJ....116.1724M,2000AJ....119.2843C} which might
point to the presence of substructure. A large fraction of the
substructures discovered so far are located in the outer regions of
our Galaxy, where the debris can remain spatially coherent for many
Gyr. This implies that it can be recovered from low dimensionality
surveys, such as 2D (using two sky coordinates), 3D (using also
distance) or 4D (including radial velocities) maps. Many surveys have
been able to discover debris using this approach, which requires
suitable tracers such as giant stars (BHB, RR Lyrae, red giant branch
(RGB)) to probe the outer Galaxy. These surveys include the Sloan
Digital Sky Survey~\citep[SDSS:][]{yanny00}, the Spaghetti Project
Survey~\citep[SPS:][]{morrison00}, the Quasar Equatorial Survey Team
(QUEST) RR Lyrae survey~\citep{vivas01}, the Two-Micron All-Sky Survey
~\citep[2MASS:][]{rocha03} and a few smaller ones.

One of the most spectacular findings has been the discovery of debris
from the Sagittarius dwarf galaxy, which spans a complete great circle
on the sky~\citep{rocha03}. We are presently witnessing the merging of
a satellite galaxy, which is just now contributing stars to build up
the outer stellar halo of our Galaxy. In fact, most of the halo
substructure discovered so far can be attributed to the Sgr
dwarf~\citep{ibata01,md01,dp01,newberg02,md04}. It is unclear whether
other accretion events in the outer halo still await discovery,
although this should probably be the case if the Galaxy was built
hierarchically. The present situation may be understood from the fact
that the streams discovered so far are dynamically the youngest, and
hence the easiest to detect. It is quite likely that other debris has
escaped detection because of its low surface
brightness. Observationally, we need to keep in mind that no surveys
have yet reached both the required depth and sky coverage.  An ideal
survey would need to have a magnitude limit not brighter than V = 19
or 20, excellent photometry to allow identification and reliable
distance estimation, preferably spectroscopic follow up to measure
radial velocities to an accuracy of $10\kms$, and its sky coverage
should allow the tracing of tidal trails over large portions of the
sky, which would require at least a few thousand square degrees.

Most of the stars in the Galactic halo are however, inside the Solar
circle. Thus, it is the inner halo that contains most of the
information related to the merging history of our Galaxy. However, in
this region of the Galaxy, the debris from a disrupted satellite
galaxy phase-mixes rapidly (a simple consequence of the shorter
orbital time scales and the more flattened nature of the Galactic
potential), erasing spatial correlations almost
completely~\citep{hw99}. Nevertheless, the debris still retains its
coherence in phase-space which is reflected in its kinematics.  For
example, if the whole stellar halo was built from merged satellites,
we would expect between 300 and 500 (mainly) cold streams in the Solar
neighbourhood, whose origin could be traced back to those
satellites~\citep{hw99,helmi03}.  

To test this prediction, large samples of halo stars with accurate 3D
velocities are required. For example, to resolve the stellar halo
velocity ellipsoid near the Sun into the individual streams at a
3-$\sigma$ level, implies that the required accuracy in any velocity
component $ \epsilon_{\rm v}$ should satisfy
\begin{equation} 
\epsilon_{\rm v}^n = \frac{\sigma_{\rm
halo}^n}{3^n N_{\rm streams}}. 
\label{eq:eps_stream}
\end{equation} 
Here $\sigma_{\rm halo}$ is the typical velocity dispersion of halo
stars near the Sun and is approximately $100\kms$, $N_{\rm streams}$
is the number of streams predicted in a small volume around the Sun,
which is of order 500, and $n$ is the dimensionality of the kinematic
survey. Thus, if radial velocities alone are available, then the
required accuracy is $\epsilon_{\rm v} \lesssim 0.07\kms$. If only the
tangential velocities are available, then $\epsilon_{\rm v} \lesssim
1.5\kms$, while in the case where all 3 components are obtainable,
$\epsilon_{\rm v} \lesssim 4.2\kms$, which is within the reach of the
RVS (for a K1III star, this precision will be achievable for stars
brighter than $V = 16.0 - 16.5$ depending on metallicity). Clearly,
knowledge of the radial velocities is critical to discover the streams
and measure their properties. It is worth emphasising that
kinematically cold streams are expected even in the full hierarchical
regime, where mergers of systems of more equal mass, rather than
simple satellite accretion, are dominant~\citep{hws03}\footnote{This
is a consequence of the conservation of phase-space density: as tidal
debris spreads out in coordinate (physical) space, the local velocity
dispersion decreases, so cold streams are expected if the mergers took
place a relatively long time ago.}. The clumpiness in the kinematics
of halo stars should thus be a distinct feature of the hierarchical
formation of our Galaxy.

Since an individual disrupted satellite may give rise to many streams
in a small volume around the Sun (for example, an object of the size
of the Small Magellanic Cloud could produce about 30 streams after
orbiting the inner Galaxy for a Hubble time, \citep[see][]{hw99}),
more sophisticated methods are needed to identify the debris
observationally. \cite{hdz00} proposed a method based on the expected
clumpiness in the space of the integrals of motion (energy, angular
momentum) for stars having a common origin. This method has been
successfully applied to the Solar neighbourhood to identify an ancient
minor merger that contributed $10$ per cent of the nearby halo
stars~\citep{helmi99}.

\begin{figure*}\resizebox{0.3\hsize}{!}{\includegraphics[height=4cm]{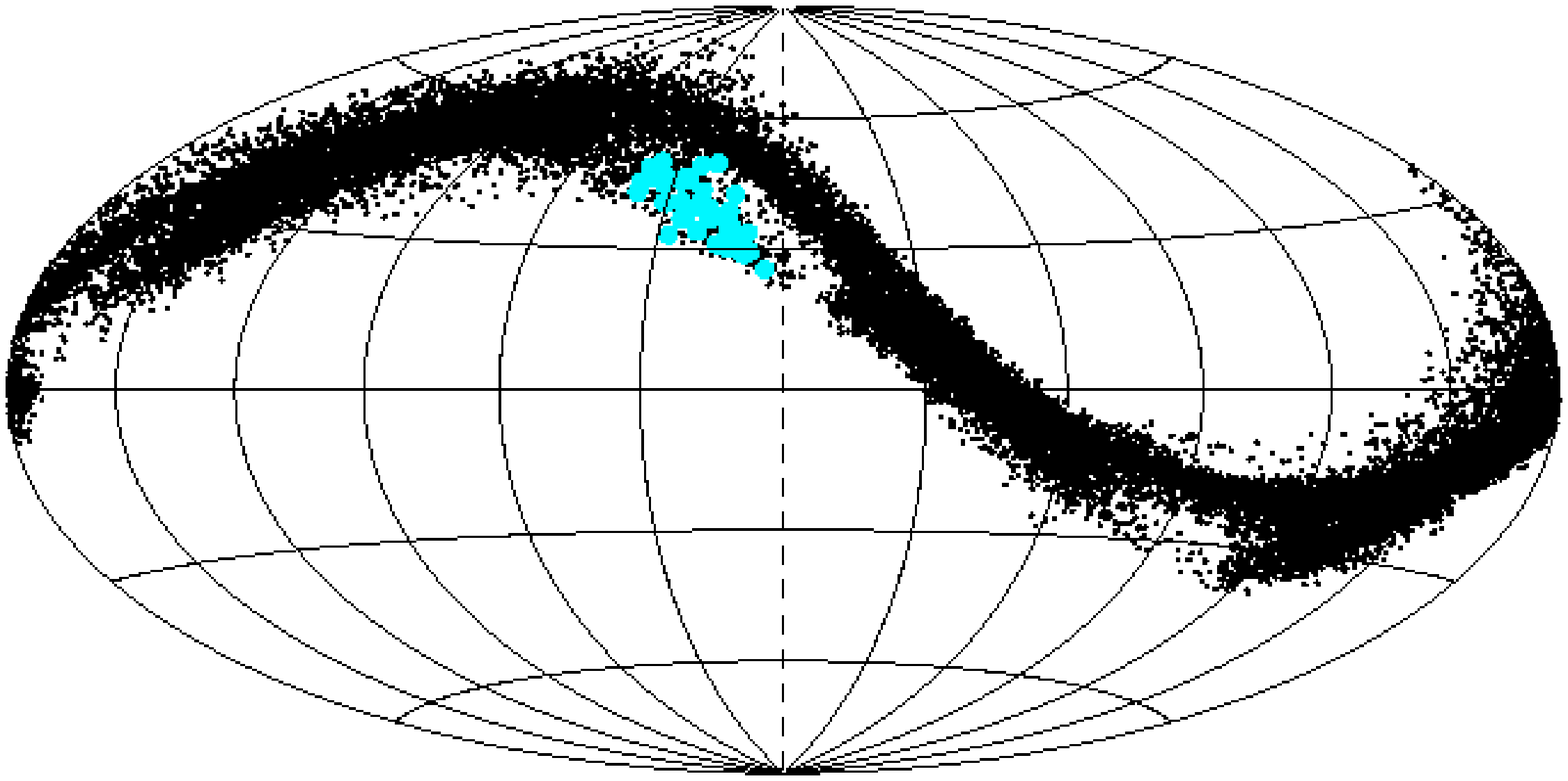}}
\resizebox{0.3\hsize}{!}{\includegraphics{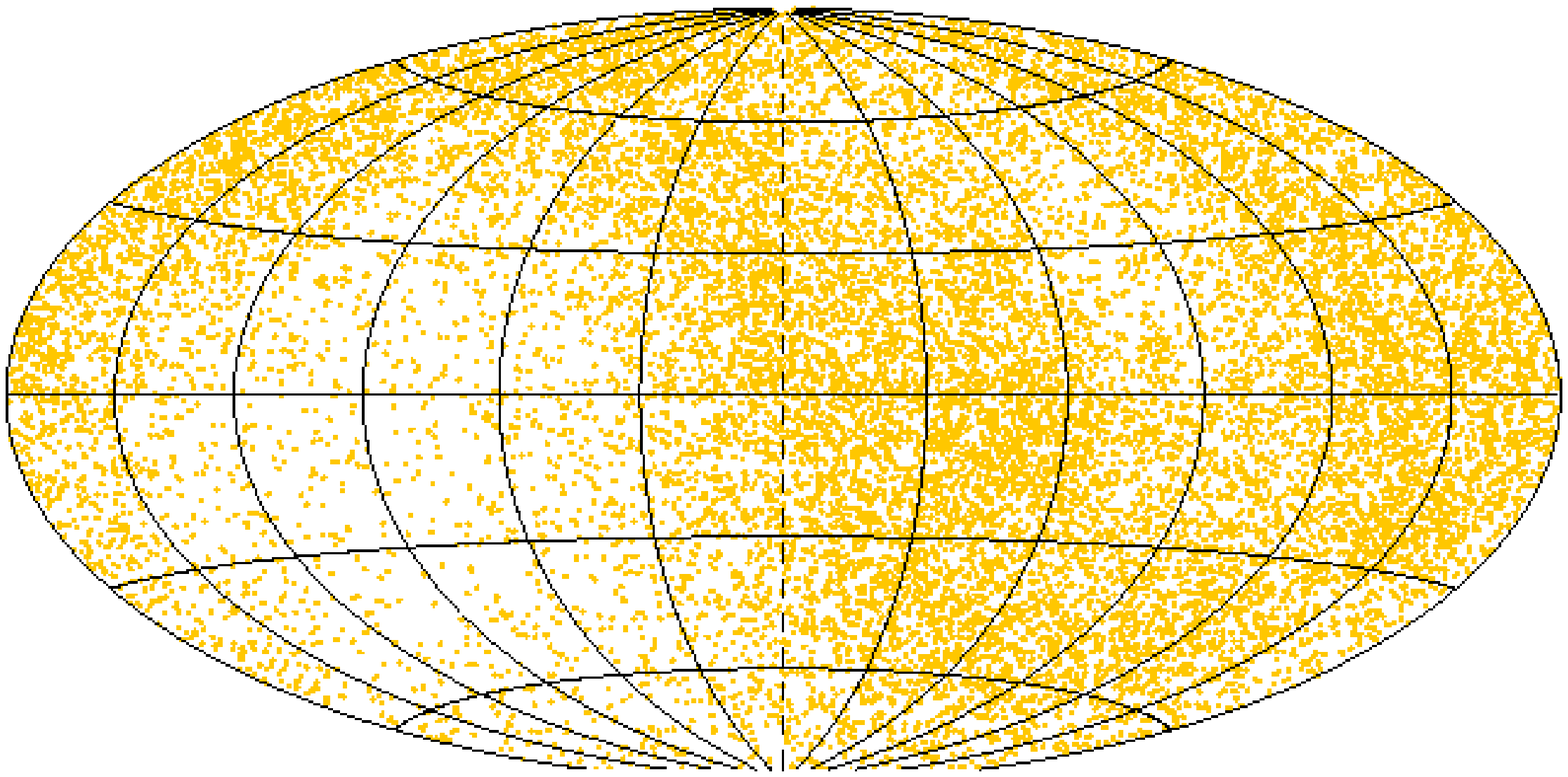}}
\hfill\resizebox{0.3\hsize}{!}{\includegraphics{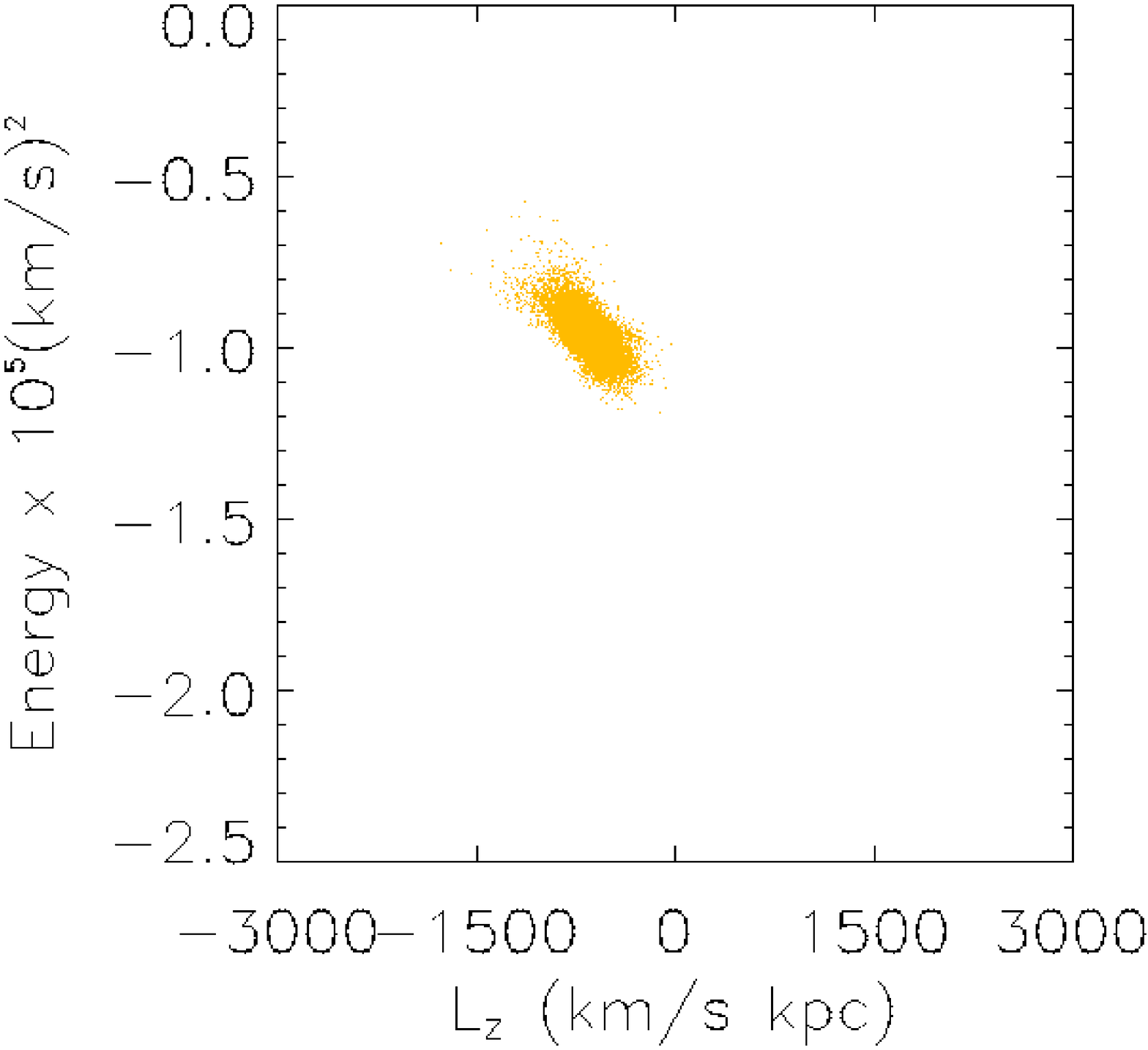}}
\resizebox{0.3\hsize}{!}{\includegraphics{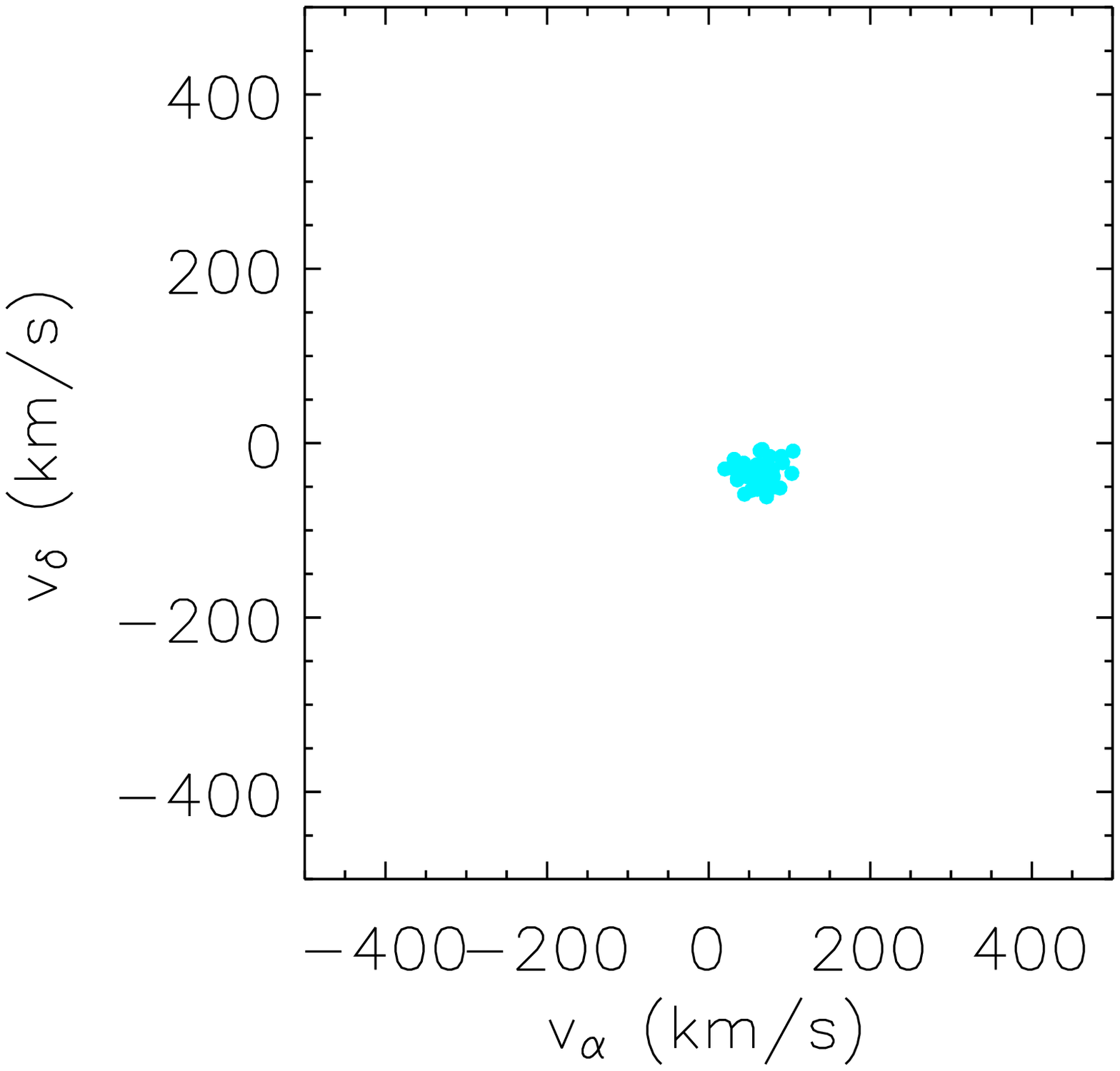}}
\resizebox{0.3\hsize}{!}{\includegraphics{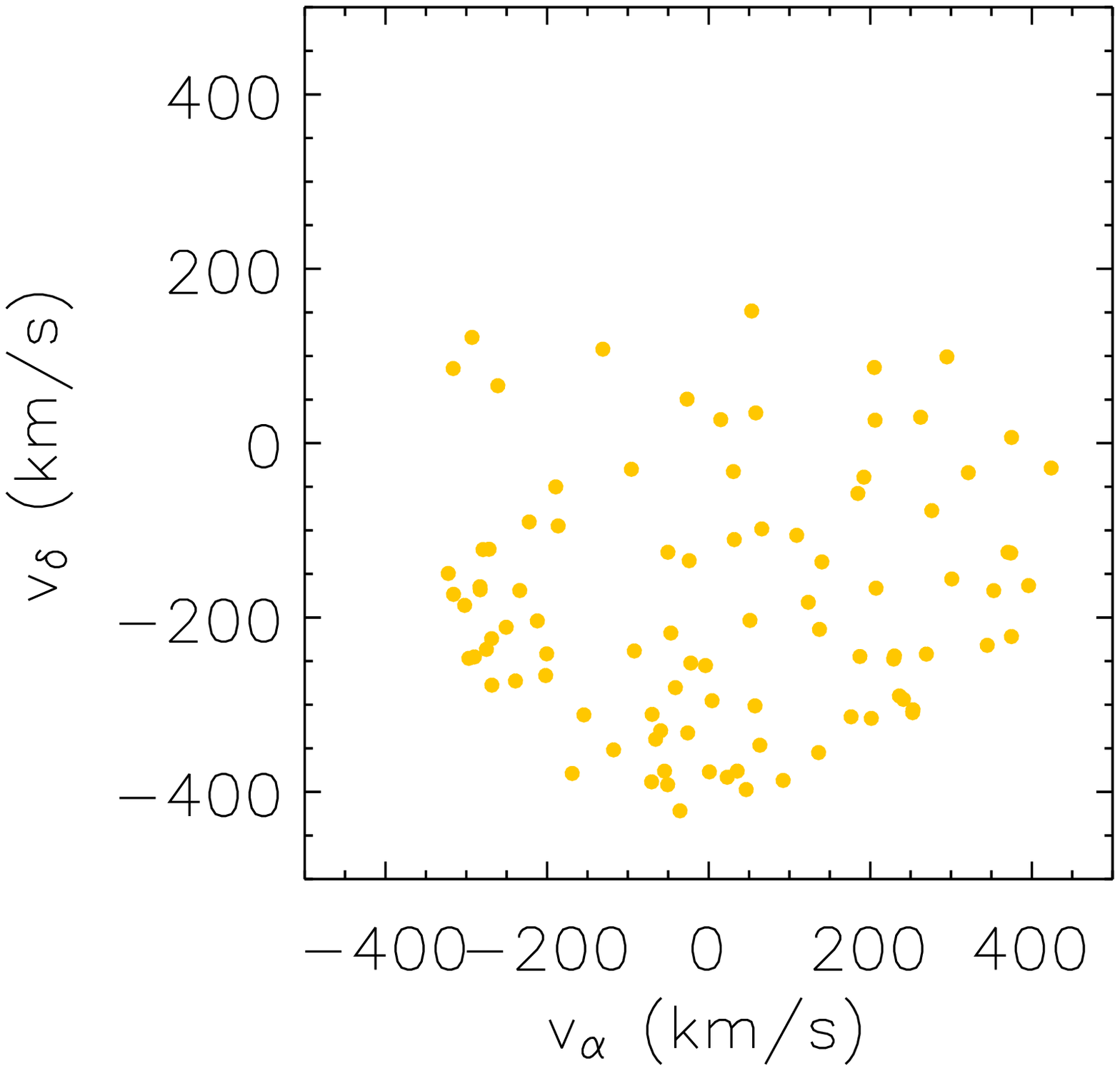}}
\hfill\resizebox{0.3\hsize}{!}{\includegraphics{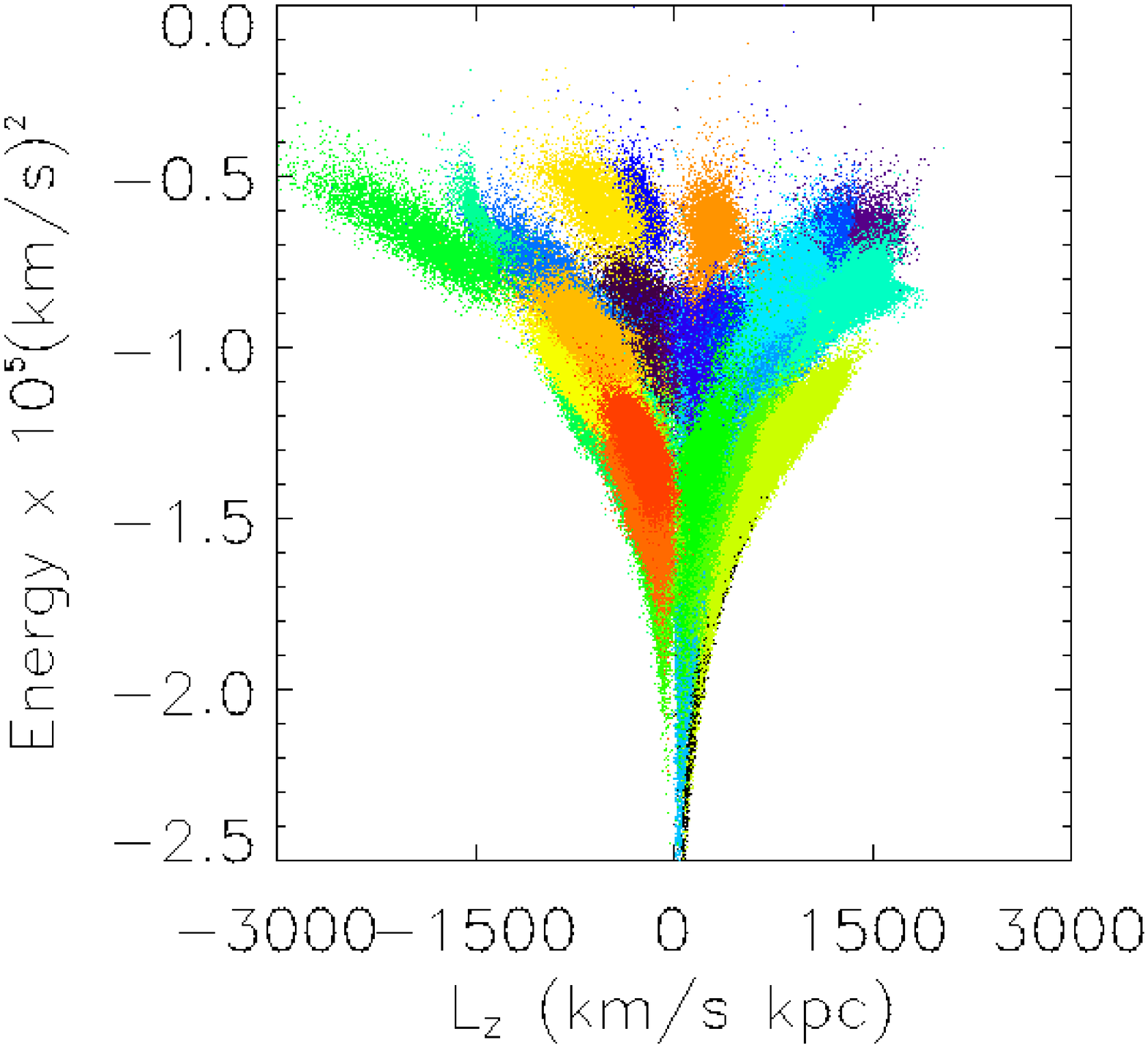}}
\caption{Distributions, in real space and velocity space, of stars
resulting from the accretion of a satellite galaxy by the Milky Way,
illustrating the importance of radial velocities in the identification
of stellar streams in the inner halo. {\bf Left:} results for a stream
in the outer halo (top panel: real space; bottom panel: velocity
space) -- the blue points represent stars within $10$ kpc of the Sun;
{\bf Middle:} results for a stream in the inner halo (top panel: real
space; bottom panel: velocity space); {\bf Right:} distribution of
integrals of motion for the inner halo satellite at the end of the
simulation (top panel) and distribution of integrals of motion for
stars originating in 33 individual satellites which could contribute
to the stellar distribution of the inner halo (bottom panel). The data
in each panel are convolved with expected {\it Gaia} observational
errors in all quantities. See text for a detailed discussion.}
\label{fig:halo_streams}
\end{figure*}

\cite{hdz00} simulated the entire stellar halo of the Galaxy starting
from disrupted satellite galaxies\footnote{Their simulations of
satellite disruption adopted a fixed Galactic potential, and hence are
quite idealistic and may not be very representative of the conditions
found in a hierarchical universe. However,~\cite{knebe05} used cold
dark matter simulations with a live Galactic potential and found that
substructures do remain quite coherent in energy-angular momentum
space.}, and ``observed'' it with the {\it Gaia} mission by convolving
the positions and velocities of the particles in the simulations with
the expected errors. They showed that the initial clumping in the
space defined by energy $E$, total angular momentum $L$ and its
$z$-component $L_z$, is maintained to a great extent even after 12 Gyr
of evolution and the error convolution. Fig.~\ref{fig:halo_streams}
presents the results of observations of the same simulations but
incorporating the most recent error estimates for the observed
parallaxes and the expected RVS radial velocity errors. In addition, a
magnitude limit of $V = 17.5$ has been assumed (\cite{hdz00} assumed a
limit of $V=15$).

Fig.~\ref{fig:halo_streams} illustrates the importance of radial
velocities in the identification of the streams in the Solar
neighbourhood. The top left-most panel in the Figure shows the sky
distribution of particles from a simulation of a disrupted satellite
orbiting for 8 Gyr in the outskirts of the Galaxy with an apogalactic
distance of about $60$ kpc and a perigalacticon of about $20$ kpc. The
particles correspond to red giant stars (each assumed to have an
absolute magnitude of M$_{\rm V}=+1$) located within 40 kpc of the Sun
(after error convolution). The region highlighted in blue contains
only those stars which are within $10$ kpc of the Sun. The panel
beneath shows their motions ($v_\alpha, v_\delta$) as would be derived
only on the basis of their proper motions and parallaxes. It is clear
that the astrometric information alone is useful for finding debris in
the outer halo.  

The middle panels in Fig.~\ref{fig:halo_streams} illustrate the case
of a stream in the inner halo. The top panel shows the sky
distribution of particles from a simulation of a satellite orbiting in
the inner Galaxy with an apogalacticon of about $12$ kpc and a
perigalacticon of about $4$ kpc. The bottom panel shows the
($v_\alpha, v_\delta$) distribution of the stars located in a sphere
of 1.5 kpc radius around the Sun and for which the velocity errors are
less than $25\kms$ and relative distance errors less than $30$ per
cent. Clearly, astrometric information alone is not sufficient to find
debris streams in the inner halo.

Given that most of the stellar halo mass is actually located in the
inner halo, it is vital that we are able to identify stellar streams
in this volume also. The top right panel of
Fig.~\ref{fig:halo_streams} shows the distribution of energy $E$ and
angular momentum about the $z$-axis $L_z$ for all the stars from the
inner halo satellite after error convolution. As ~\cite{hdz00} found,
the stars retain their clustering in the space defined by the
integrals of motion. For each star in the RVS sample, both $E$ and
$L_z$ can be computed. The bottom right panel shows the ($L_z, E$)
distribution for stars originating in 33 different satellites that
could be populating the present day stellar halo. For each star $E$
and $L_z$ have been computed after proper error convolution on all
observed quantities. A standard friends-of-friends algorithm applied
to the ($L_z, L, E$) space (where $L$ is the magnitude of the total
angular momentum vector) is able to recover two thirds of all the
satellites. The use of 6-D information is critical for the recovery of
these accretion events: this would not be possible without the radial
velocity information provided by the RVS. 

 More realistic, cosmologically motivated, models of the Galaxy and in
particular of its stellar halo, as well as a deeper understanding of
the phase-space structure of the debris from mergers, need to be
produced. These would enable us to develop more refined methods to
recover the debris observationally and, once the {\it Gaia} data are
available, to unravel the formation history of our Galaxy.

\subsection{Dark Matter in the Milky Way}
\label{sec:dark_matter}

Our knowledge of the total mass and extent of the dark halo of the
Milky Way is sadly lacking given its importance for comparisons with
theories of galaxy formation~\citep[e.g.][]{kochanek96,we99}. The RVS
will significantly enhance our understanding of the dark matter
distribution within 50 kpc of the Galactic centre.

The volume and column mass density of the Galactic disc place limits
on the nature of the dark matter which makes up the halo. Locally, we
can place robust limits on the mass in stars and gas based on star
counts and our knowledge of the stellar mass function. The total mass
implied by models of the stellar kinematics can be compared with the
observed mass~\citep[e.g.][]{kg89,ccbp98,clmva03}. Current estimates
suggest that the two approaches yield similar values, which argues
against the presence of a significant amount of dark matter in the
Galactic plane and therefore the existence of dissipational forms of
dark matter. The scale-height of any unseen component can be inferred
from observations of populations at a range of radii and heights above
the Galactic plane which will show the variation of volume density
within the disc. As~\cite{kg89} discuss, a key uncertainty in current
estimates of the local mass density lies in the uncertain variation in
the ``tilt'' of the stellar velocity ellipsoid as a function of height
or, in other words, whether the velocity ellipsoid is aligned with the
axes of a spherical (maximal tilt) or cylindrical (no tilt) coordinate
frame. The accuracy of the RVS velocities will be sufficient to map
out the velocity structure over a large volume surrounding the
Sun. For example, the radial velocities of metal poor K1III stars will
be measurable with accuracies of $< 5\kms$ to a limiting magnitude of
${\rm V}\simeq16$, corresponding to a distance of $10$ kpc (assuming
M$_{\rm V}=1$). Radial velocities are necessary to determine the
detailed shape of the velocity distribution in a model-independent
way.

Another important quantity related to the mass of the Milky Way is the
local escape speed $v_{\rm e}$. An estimate of $v_{\rm e}$ may be
obtained by studying the high velocity tail of the stellar velocity
distribution in the Solar
neighbourhood~\citep[e.g.][]{lt90,kochanek96,meillon99}. Current
samples include only a few tens of high velocity
stars~\citep{clla94,meillon99,sakamoto03} which only weakly constrain
the shape of the velocity distribution. Radial velocities are
necessary for this work as any underestimate of the space velocities
of the stars translates directly into an underestimate of the escape
velocity. Further, given that the estimate of $v_{\rm e}$ currently
requires assumptions to be made about the shape of the velocity
distribution of weakly bound stars, large samples of high velocity
stars with accurate proper motions and radial velocities are essential
to specify the shape without recourse to models. The accuracies
required are not particularly demanding as the typical high velocity
stars are moving at velocities in excess of $350\kms$. BHB stars
belonging to the stellar halo are a valuable tracer population for
this work as they are numerous and relatively bright (M$_{\rm V} \sim
0.6$). For a typical metal-poor, halo BHB star, the RVS will be able
to obtain velocities to a limiting magnitude of about 16.5,
corresponding to distances of about $15$ kpc. There will be about
10\,000 such stars in the observable volume, which will place strong
constraints on the actual shape of the local halo velocity
distribution.

The question of the full extent and mass of the Milky Way halo remains
an open issue. Recent estimates~\citep[e.g.][]{kochanek96,we99} which
include all available data on tracer objects outside $20$ kpc still
have uncomfortably large error bars. A radial velocity accuracy of
$10-15 \kms$ is sufficient for this work. Stars at the tip of the RGB
will be observable to distances of about $50$ kpc while asymptotic
giant branch (AGB) stars (M$_{\rm V} \sim -2.5$) will be observable
out to about $60$ kpc. Samples of several hundred of each of these
tracers will be observed by the RVS out to the orbit of the LMC. In
addition, CH-type carbon stars with M$_{\rm V} \sim -2.5$ will be
observable out to $60$ kpc. The density of these stars on the sky is
about four times higher than that of AGB
stars~\citep{1998MNRAS.294....1T} and so a sample of about 1000
objects is expected.

Simulations show that a sample of 500 tracers extending to large radii
can reduce the error on the estimate of the enclosed mass to about
$20$ per cent and remove systematic errors~\citep[see Fig.~12
of][]{we99}. The RVS will provide such a sample extending to about
$50$ kpc.  More importantly, it will also be possible to look for
correlations in the motions of the tracers which would provide
information about the origins of the stellar halo.  Recent
observations of stars in the outer halo suggest that a significant
fraction of this Galactic component may be composed of
streams~\citep{majewski03}. Knowledge of such correlated motions is
important for estimating the mass of the Galaxy as mass estimators
generally assume that all tracers are drawn independently from an
underlying velocity distribution. The unambiguous characterisation of
a stream requires all three components of the stellar velocity --
radial velocities alone place much weaker constraints, except for
stars near the turning points of an orbit~\citep[e.g.][]{clewley05}.
{\it Gaia} parallaxes will not provide accurate distances for these
distant halo tracers. However, the {\it Gaia}-calibrated distribution
of absolute magnitudes for particular tracers, including any
dependence of absolute magnitude on metallicity or other internal
parameters, will make it possible to obtain reliable photometric
distances. If we conservatively assume that these distances can be
calculated with an accuracy of about $10$ per cent, transverse
velocities for all the above halo tracers can be obtained with
uncertainties of about $10-15$km\,s$^{-1}$. Thus, we can expect to
have full space motions for samples of several thousand stellar
tracers out to $50-60$ kpc. Obtaining all-sky radial velocity coverage
to an equivalent magnitude limit from a ground-based survey would be
extremely time-consuming.

The nature of the dark matter which makes up the dark haloes of
galaxies is also an open question. In addition to improving our
knowledge of the distribution of mass within the Milky Way which
itself constrains the properties of the dark matter, the RVS will also
provide a direct test of the lumpiness of the dark halo. ~\cite{ycg04}
investigated the expected distribution of wide stellar binaries in
haloes composed of massive compact objects (MACHOs) and found that
encounters with MACHOs tend to disrupt the widest binaries and lead to
a cut-off in the distribution of binary separations. Using the
observed distribution of binary separations from a large sample of
halo binaries~\citep{cg04}, \citeauthor{ycg04} conclude that the dark
halo cannot contain a significant fraction of its mass in the form of
compact objects with masses above $43$M$_\odot$. The velocities
furnished by the RVS, in combination with the {\it Gaia} proper
motions, will permit confirmation of the physical association of all
the binaries which \citeauthor{ycg04} consider by providing the full
space motions of their components. Further, it will be possible to
investigate the Galactic orbits of these binaries which has
implications for their expected survival times in a lumpy halo. In
addition, a vastly larger sample of wide halo binaries will be
obtained which will make it possible to apply this test of the nature
of the dark matter over an increased volume of the dark halo.

\subsection{Dis-entangling the thin and thick disc populations}
The study of the kinematics of the disc of the Milky Way is important
for a number of reasons. First, the thin and thick discs are major
components of the Galaxy and provide an opportunity to study the
internal dynamics of a galactic disc at a level of accuracy which is
impossible for external galaxies. Secondly, information about the
formation processes which produced the disc can be inferred from the
motions of its stars. The RVS will be an essential tool in the study
of the disc as it will provide velocities of sufficient accuracy to
probe all aspects of disc kinematics.

In order to investigate the formation of the Galactic thin and thick
discs, it is vital to obtain precise information on the global
Galactic kinematic properties (i.e. the velocity ellipsoids) so that
deviations from expected behaviour can be reliably derived.  At
present, the kinematics of thin disc stars are poorly known: the
structure of the velocity ellipsoid, its vertex deviation and
inclination with respect to the Galactic Plane rest on a small sample
of local stars. The vertex deviation measures the extent to which the
local velocity ellipsoid is radially aligned and is strongly affected
by the presence of moving groups in the local sample (see
\S~\ref{sec:movinggroups}). \cite{2003A&A...398..141S} suggest vertex
deviations which range from about $25$ degrees for young stars to
about $0$ degrees for older stars, while~\cite{dehnenBinney98} find
that older stars have a vertex deviation of about $10$ degrees.
Additionally, to date, the vertical tilt parameter $\lambda(R)$ of the
thin disc velocity ellipsoid remains ill-determined.  The value of
$\lambda$ indicates whether the velocity ellipsoids are aligned with
the coordinate axes of a spherical ($\lambda = 1$) or cylindrical
($\lambda = 0$) coordinate system. In the first case the velocity
ellipsoid points towards the Galactic center, while in the second it
is everywhere parallel to the Galactic plane. The value of $\lambda$
is, in fact, strongly related to the coupling of the U and W
velocities\footnote{The U direction is defined to be positive in the
direction of the Galactic centre, while the W direction is positive
towards the North Galactic Pole.} and can only be properly constrained
using three-dimensional velocities.  It is found to vary between $0.4$
and $0.7$ at the Solar circle \citep{1991MNRAS.253..427C,
1999A&A...341...86B}, while no information is available at more
distant locations. Although the value of $\lambda$ is directly related
to the shape of the gravitational potential, the current range of
values are consistent with a wide range of mass distributions
including an exponential disc mass distribution embedded in either a
spherical halo with a flat rotation curve ($\lambda=0.7$) or a prolate
halo ($\lambda<0.5$). More precise determinations of $\lambda$,
including its variation with position in the disc, are required and
will be provided by {\it Gaia}.

In addition to the alignment of the velocity ellipsoid, the magnitudes
of the components of the velocity dispersion tensor may also vary
according to the age of the stellar population considered.
\cite{1989AJ.....97..139L} argue that a gradient in velocity
dispersion as a function of stellar age ought to exist. The
age-velocity dispersion relationship is based on samples of stars in
the Solar vicinity. For example, \cite{2001gddg.conf...87Q} find that
the heating of the disc saturates after about 2-3 Gyr. However, recent
results by \cite{2004A&A...418..989N} suggest a continued heating of
the disc. The question is still under debate and larger data sets are
required in order to resolve this issue.

Information about the formation processes which produced the thin and
thick discs can also be gleaned from the rotational character of the
discs.  The presence, if any, of vertical velocity gradients in the
stellar components might be considered as another relict of the
formation process \citep{2000AJ....119.2843C, 2003A&A...398..141S,
gwn02, 1995ApJS...96..175B,freeman02}.  $N$-body simulations have
demonstrated that radial mixing (e.g. due to the presence of transient
spiral arms) can wash out chemical and age gradients close to the
Galactic plane on time scales of a few Gyr~\citep{sellbinney}. Thus,
it is possible that only weak gradients will be detectable in the thin
disc population. However, the thick disc should preserve a memory of
its formation process since radial mixing is not so effective outside
the Galactic plane and vertical gradients in metallicity and velocity
dispersion can, in principle, trace the formation history. These
gradients are currently very poorly constrained. On the one hand,
velocity gradients perpendicular to the Galactic Plane are excluded by
\cite{2003A&A...398..141S}; on the other hand, it has been suggested
that the velocity dispersion and rotational velocity both decrease far
from the disc, with a velocity gradient of about 30
km\,s$^{-1}$\,kpc$^{-1}$ \citep{2000AJ....119.2843C}. The RVS will
make a substantial contribution to the resolution of this issue by
providing velocities for a large sample of tracers covering a
significant fraction of the low density regions of the thick disc. For
example, the velocities of K1III giants (with M$_{\rm V}=1$) will be
measured to an accuracy of about $25\kms$ up to distances of $25$ kpc
from the Sun. Preliminary simulations show that \textit{Gaia} will be
able to detect the presence of a gradient in the thick disc rotation
velocity of the order of $10\kms$\,kpc$^{-1}$~\citep{bertelli03}.

It is also likely that the disc components of our Galaxy contain
debris from past minor mergers. In particular, the thick disc could be
the product of the heating of an ancient thin disc by a relatively
large satellite~\citep[e.g.][]{quinn93,vw99,freeman02}. Another
related possibility is that the thick disc is constituted only by
debris from disrupted satellites having initially low inclination
orbits~\citep{anse03}: simulations show that a substantial fraction of
the old disc stars could have formed in external systems. Finally, a
multi-component structure with tracers of both mergers and accretions
has been suggested by \cite{gwn02}. In all cases, kinematic
substructures are expected to remain in the disc and these should be
detectable by the RVS, as well as in large radial velocity surveys
such as the RAdial Velocity
Experiment~\citep[RAVE:][]{2003gsst.conf..381S}, although the latter
will survey a much small fraction of the disc than that covered by the
RVS.

Perhaps the earliest example of such an old kinematic structure with
thick disc kinematics is the Arcturus group~\citep{eggen71}, whose
nature needs to be confirmed~\citep{nhf04}.  The recent survey of
\cite{gwn02} probed more distant regions of the thick disc, and found
other hints of substructure. However, without full phase-space
coverage for stars over a large fraction of the disc it remains
difficult to make sense of the different moving groups, and to
establish whether or not they have a common progenitor.

The outer disc of the Galaxy is very complicated, with observational
evidence both for the warping and flaring of the stellar distribution
outside the Solar circle~\citep[e.g.][]{ds89,egtb98}. Kinematic
studies to date present a confusing picture of the dynamics of the
warp~\citep[e.g.][]{mys88,sdlb98}. The origin of the warp is also
unclear with suggested possibilities including interactions between
the disc and either the halo, an infalling satellite or infalling
gas~\citep[see][ for a discussion]{rrdp03}. While stellar proper
motions are most suited for studies of warp kinematics, the RVS radial
velocities can detect the bulk motion associated with the warp. For
example, \cite{feitspick87} discuss the foreshortening effect of the
warp on the distribution of gas velocities in the Galaxy. For a star
at $\ell=90^\circ$ (the direction of the maximum height of the warp
above the Galactic plane) moving on a circular orbit, the
foreshortening effect amounts to a change in the radial velocity of
the star of about $4\kms$. The line of sight component of the rotation
about the nodal line of the warp is also comparable in
magnitude~\citep{msy93}.

\begin{figure}
\resizebox{\hsize}{!}{\includegraphics[height=4cm]{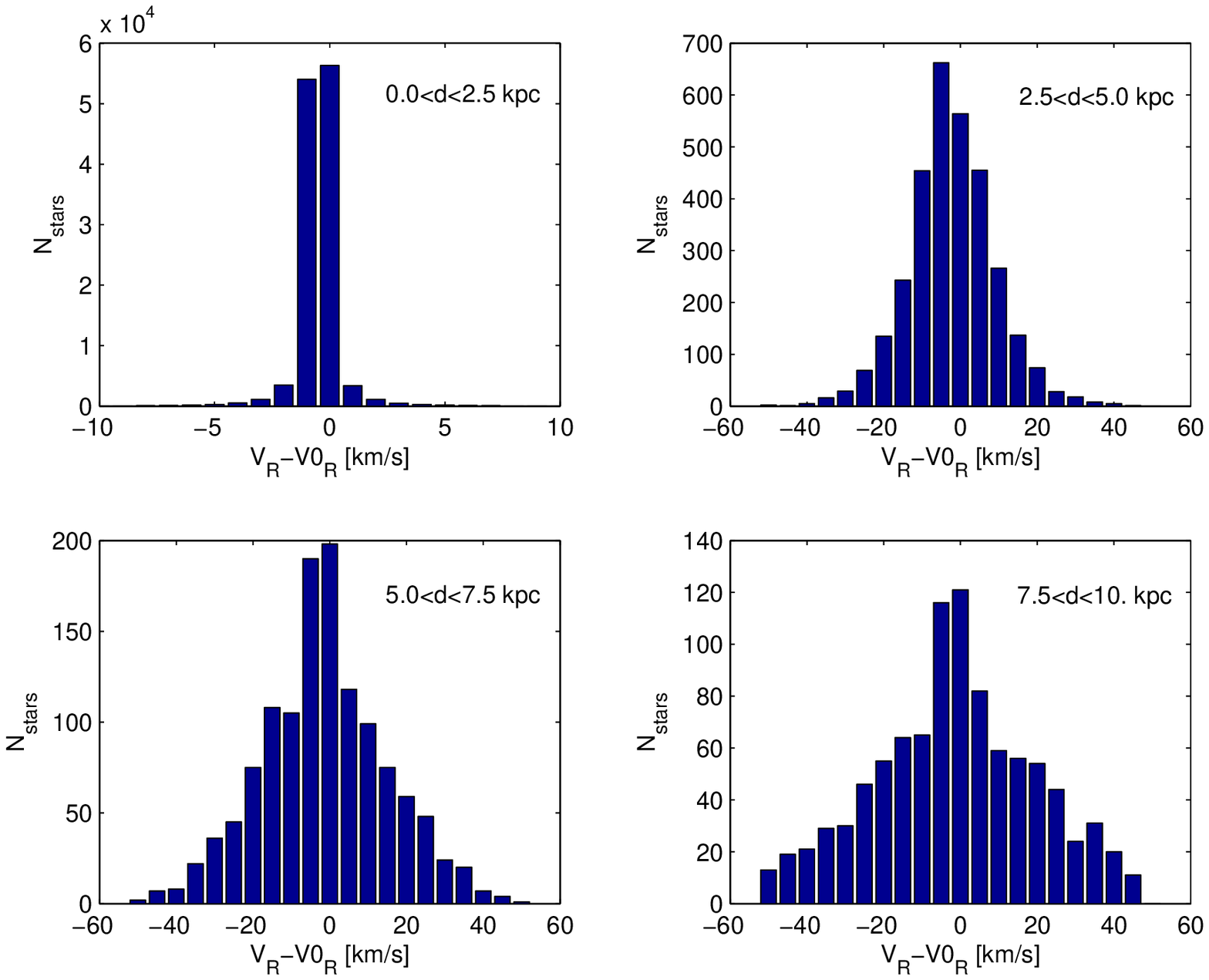}}\\
\resizebox{\hsize}{!}{\includegraphics[height=4cm]{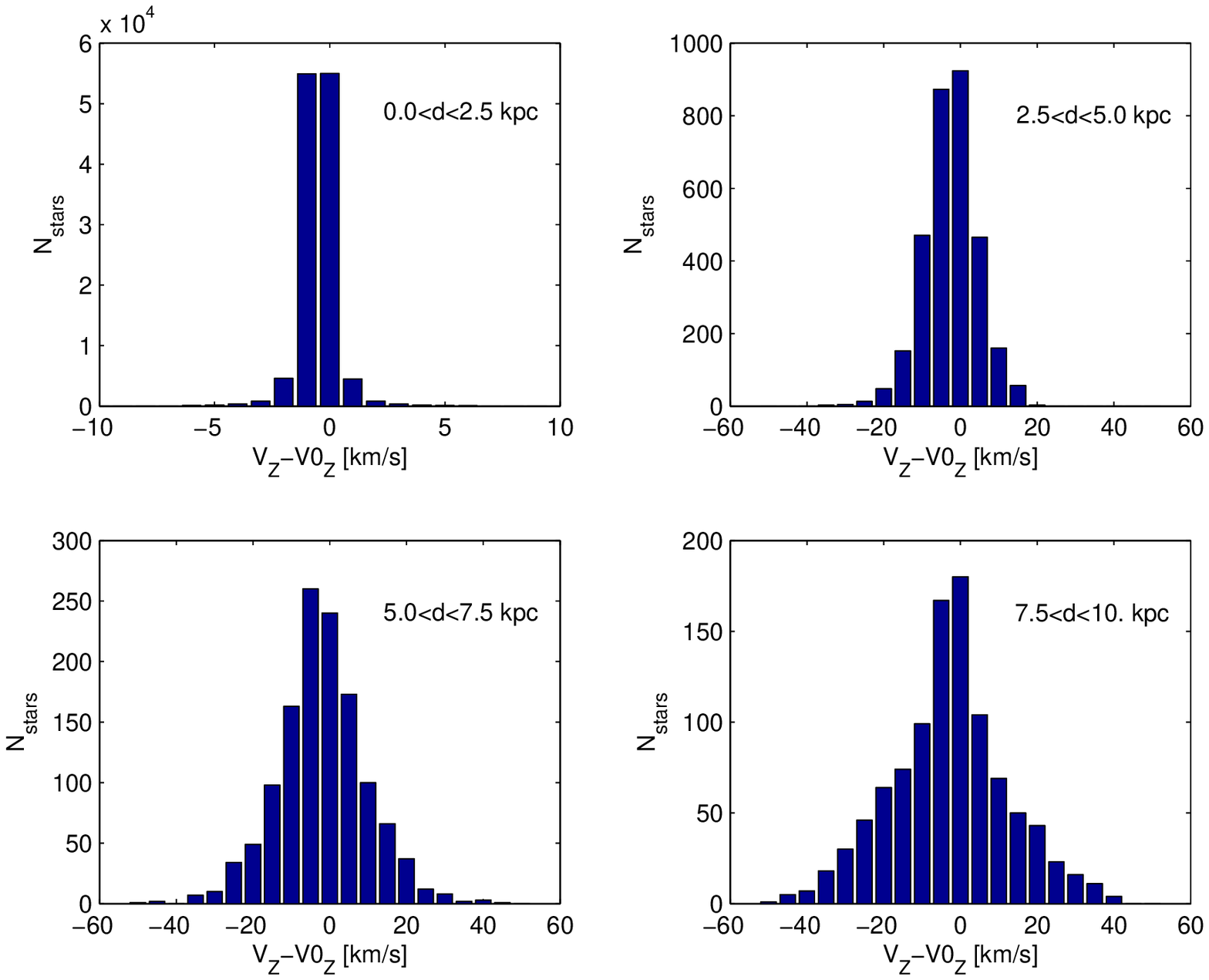}}\\
\resizebox{\hsize}{!}{\includegraphics[height=4cm]{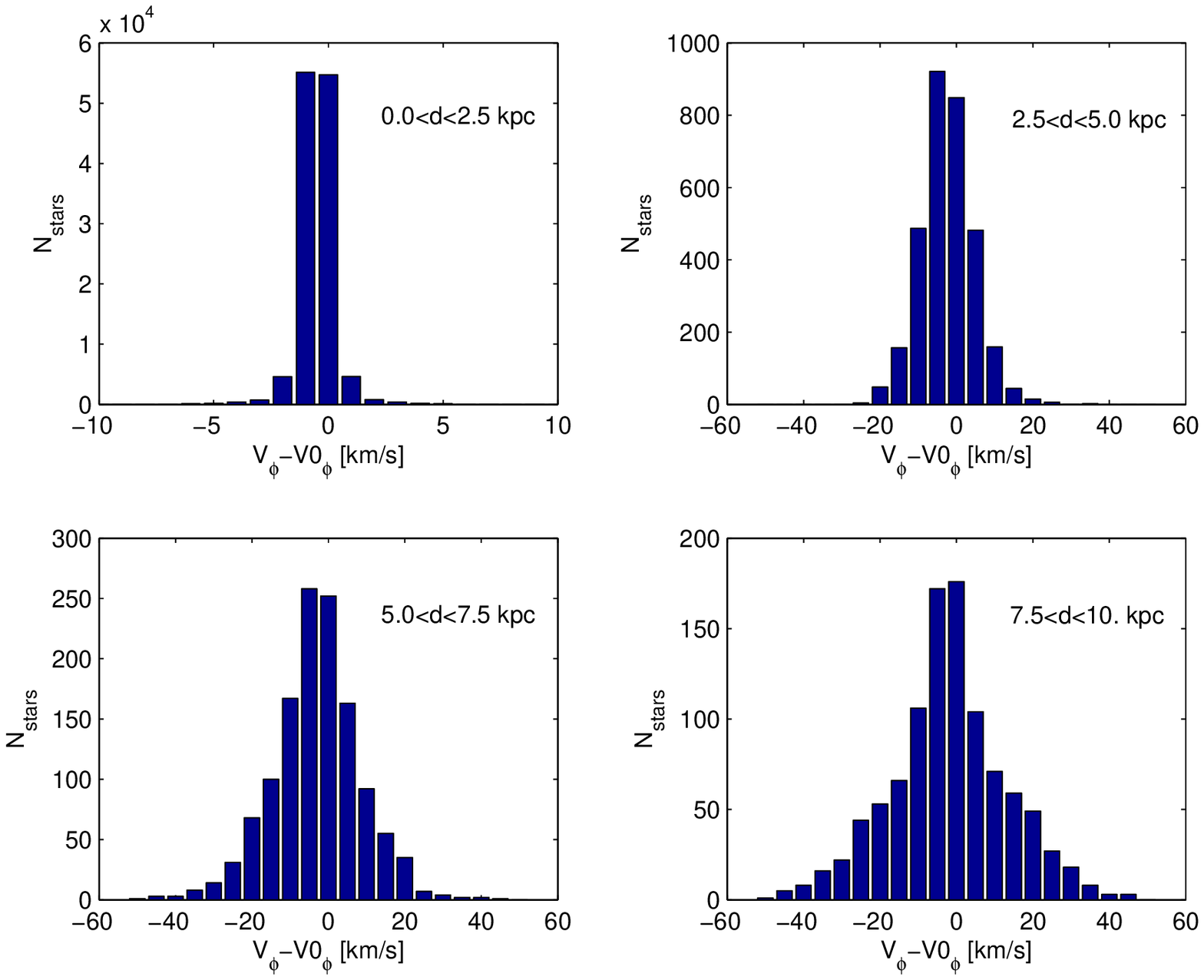}}
\caption{Distribution of velocity residuals at various line-of-sight
distances in the direction $(l,b)=(270,45)$ when {\it Gaia} accuracies
on proper motions, radial velocities and distances are taken into
account. The distributions were calculated from a Monte Carlo
simulation of about 10\,000 stars of thin disc, thick disc and halo
populations brighter than $V=17$. The three components of the observed
velocity (radial, $v_{\rm R}$; vertical, $v_{\rm z}$; azimuthal,
$v_\phi$) are shown separately.}
\label{errors}
\end{figure}
Fig.~\ref{errors} presents the results of a Monte Carlo simulation of
about 10\,000 stars of thin-disc, thick disc and halo populations
brighter than $V=17$ at $(l,b)=(270,45)$ derived using the Padova
Galaxy Model~\citep{bertelli95,vallenari03,bertelli03} including
expected {\it Gaia} accuracies on proper motions, parallaxes and
radial velocities\footnote{{\it Gaia} is expected to measure
parallaxes with a precision $\sigma_\pi/\pi < 10$ per cent for a
bright star having M$_{\rm V}=0$ up to distances of $10$ kpc and with
a precision of 1 per cent up to $2.7$ kpc. Proper motions will be
known with a precision of a few $\mu$as/yr for stars brighter than
${\rm V}=14-15$ and of about $30\mu$as/yr for stars of ${\rm V}=18$,
depending on the spectral type.}. The differences between the observed
and true velocity are plotted. The errors increase with distance,
being less than $1\kms$ closer than 1 kpc, $10\kms$ at 4 kpc, and
$20-30\kms$ at 10 kpc. This accuracy should be compared with the
typical velocity dispersion of the thin disc (less than $10\kms$),
thick disc ($40-70\kms$) and the halo ($100-200\kms$),
respectively. Thus, it is clear that the velocity ellipsoids of both
the thin and thick discs, as well as all other kinematic
characteristics of their populations, can be derived with
unprecedented accuracy.

The spiral arms associated with spiral density waves induce systematic
motions of stars and gas both along and across the arms themselves and
in the inter-arm regions which can be used to map out the
gravitational potential of the arm~\citep{lin69}. Observations of OB
associations in the Cygnus-Orion arm have shown that the magnitude of
these systematic velocities is between $10$ and
$20\kms$~\citep{sitnik}. \cite{feitspick87} found similar velocities
in their study of simulated gas velocity fields in the disc. Thus,
individual radial velocities with accuracies of $5\kms$ are sufficient
to map out the velocity field associated with a spiral arm. The RVS
will be able to obtain sufficiently accurate velocities of B5V stars
to distances of about $2.5$ kpc and of Cepheid tracers to distances of
$6-10$ kpc. These data will facilitate detailed comparisons with the
spiral arms seen in numerical simulations and may provide clues to the
true origin of spiral structure which is currently
unclear~\citep{sellwood00}.

\subsection{Moving groups}
\label{sec:movinggroups}
It is now well established that the distribution of stellar velocities
in the Solar neighbourhood is not smooth -- well-defined groupings in
velocity space are clearly visible, for example in the proper motion
data from {\it Hipparcos}~\citep[e.g.][]{famaey05,Dehnen98}. The
clumping is not random as similar features are seen in the velocity
distributions of stars of different colours. The similarity between
the colour-magnitude diagrams (CMDs) derived from stars which are
members of moving groups and those of open clusters led to the
suggestion that these groups were the remnants of dissolved clusters
whose motion remains correlated long after the spatial coherence of
the cluster has disappeared. However, the wide age-range observed for
stars in certain moving groups is difficult to understand in this
picture, suggesting that at least some groups must have had a
different origin. Recent work by \cite{swt04} suggests that certain
old moving groups may be the product of disc heating by temporal
variations in the disc potential due to the passage of stochastic
spiral density waves. This would explain why stars of a variety of
ages may be found in a single velocity structure. Outward moving
streams (e.g. the Hercules stream) may be the result of chaotic
relaxation of stars in the non-regular parts of phase space created by
the central bar~\citep[e.g.][]{fux01}. In addition, the outer Lindblad
resonance of the bar is near the Solar radius, which may give rise to
some of the observed velocity features~\citep{Dehnen00}.

The characterisation of moving groups is of great interest as it
provides clues to many aspects of the internal structure of the Milky
Way disc. The presence of these groups complicates the determination
of the global properties of the disc populations. For example, the
inclusion of a moving group in the local velocity sample can mimic the
presence of vertex deviation of the velocity ellipsoid. Even the
determination of the solar motion relative to the local standard of
rest has been shown to depend sensitively on whether or not all the
nearby moving groups are included in the stellar sample used to
measure this motion~\citep{famaey05}. As was discussed above, the
complete characterisation of the velocity distribution of disc stars
is an essential step towards the development of a full understanding
of the origin and evolution of the Milky Way disc. It will be greatly
facilitated by the availability of radial velocities: the
determination of the local velocity distribution using proper motions
alone is equivalent to performing a difficult
de-projection~\citep{Dehnen98}. The addition of the RVS radial
velocities will make it possible to determine the full
(i.e. unprojected) velocity distribution function for known
groups. The identification of new groups will be made more robust
through the use of integrals of motion to define membership rather
than merely clumping in velocity space. The velocity accuracy required
is modest: individual velocity errors of about $5 \kms$ will suffice
to distinguish between typical moving groups. As Fig.~\ref{errors}
shows, moving groups will be easily identified out to distances of
around $2.5$ kpc.

\subsection{OB Associations}
In recent years, observations of star forming regions have shown that
stars almost never form in isolation -- most, if not all, stars form
in groups or clusters~\citep[e.g.][]{LEDG91}. Thus, a full
understanding of the details of star formation requires robust
observations of young stellar clusters. The Milky Way disc contains a
large number of clusters on a range of spatial scales from the low
density, unbound and expanding OB associations (e.g.  Perseus OB2) to
the massive open clusters (e.g. the Hyades and Pleiades clusters). The
{\it Gaia} mission, and the RVS in particular, will improve our
understanding of these objects by allowing the accurate determination
of association/cluster membership based on distance and
three-dimensional kinematics. The RVS will also facilitate
measurements of the internal motions of stars within these systems. In
addition, their collective space motions within the Galactic potential
will yield insights into the connection between star clusters and the
main stellar populations of the disc.

OB associations are unbound collections of recently formed stars which
allow us to probe the initial mass function, primordial binary
fraction, and early-time dynamics resulting from the process of star
formation.  They are ideal locations in which to determine the
fraction of stars which form in binaries as they are young stellar
systems in which dynamical effects have had relatively little impact
on the primordial binary population~\citep[e.g.][]{brown01}. Their
internal velocity dispersions are typically a few $\kms$. Membership
has traditionally been determined using the convergent point method
(which uses only positions and proper motions but not parallaxes)
although recent studies have introduced modifications to this
approach~\citep{bruijne99} as well as new techniques such as the
``spaghetti method''~\citep{ha99,ah01} which take account of parallax
information. 

It has long been known that the intrinsic expansion of OB associations
resulting from gas loss cannot be determined using proper motion data
alone~\citep{blaauw64}. In particular, it is impossible to distinguish
between a radially expanding group at rest with respect to the
observer and a cluster approaching along the line of sight.
As~\cite{perryman98} discuss, the standard convergent point method for
the identification of members relies on assumptions (small internal
velocity dispersion, no internal kinematic structure or rotation)
regarding the space motion of the cluster which can only be tested
once radial velocities are available. For example, a net rotation may
be inferred for a cluster with a large internal velocity dispersion if
proper motions alone are used to determine membership due to the fact
that the proper motion selection will produce an artificial flattening
of the cluster in velocity space. The presence or absence of rotation
in star forming regions may provide important clues to the origin of
the angular momentum observed in some globular
clusters~\citep[e.g.][]{meylanheggie96,vl00,AndKing03} and it is
therefore important to be able to determine whether or not apparent
rotation is real. The accurate determination of the orbits of young
clusters and associations is of importance for the investigation of
their relation to the surrounding field star population. In this
context, the decomposition of any apparent radial velocity into bulk
motion and expansion is clearly important. Only radial velocities,
with accuracies of a few $\kms$, can resolve the situation.

Radial velocities are an essential complement to proper motions in the
correct identification of members of OB associations which is the
first step towards their use in discussions of the issues raised
above. For example, using radial velocities with errors $\sim
3\kms$,~\cite{SBHdZ03} have shown that in the case of the association
Perseus OB2, a number of interlopers were included in the list of
possible members determined by~\cite{dZHBBB99} based on {\it
Hipparcos} parallaxes and proper motions alone. The inclusion of
radial velocity information in the analysis provided an additional
selection criterion which made it possible to identify these
interlopers as unrelated field stars, due to the $10\kms$ offset in
radial velocity between the association and the disc stars along the
line of sight. More accurate assessment of membership probabilities
for stars near associations is important as it allows uncontaminated
CMDs to be plotted which in turn provide important information about
the initial mass function.

Associations are often studied using only their brighter, early type
members (the OB stars). The intrinsic difficulty of obtaining radial
velocities for early type stars, particularly when only the spectral
region around the CaII triplet is available, is well known. In the
case of the RVS, for a slowly rotating B5V star ($v\sin i = 50\kms$)
with absolute magnitude $M_{\rm V}=-1$, the expected velocity errors
are about $2.5\kms$ at $1.5$ kpc ($V=10$) and about $5\kms$ at $2.5$
kpc ($V=11$), in the absence of reddening. Although radial velocities
at this level of accuracy are insufficient to investigate the internal
dynamics of associations (due to their small intrinsic velocity
dispersions), they will be useful for the determination of association
membership. In addition, spectral information allows the estimation of
the extinction towards individual stars which gives a further clue to
whether or not a given star should be associated with a particular
cluster.

Associations also contain large numbers of lower mass ($M \lesssim
2$M$_\odot$) stars~\citep[e.g.][]{brown01} for which the RVS will
easily furnish radial velocities at the $1-3\kms$ accuracy level. The
exact numbers in particular associations will depend strongly on the
local levels of extinction. However, we can estimate the range of
masses which the RVS will be able to study using the properties of two
known associations. The nearest OB association is Scorpius OB2, which
has a distance modulus of $5.3-5.8$ magnitudes~\citep[equivalent to
distances of $116-144$ pc;][]{dZHBBB99}, solar
metallicity~\citep{eggen98}, an age of 5-15 Myr~\citep{degeus89} and
visual extinction in the range A$_{\rm V} = 0.1-1.3$
magnitudes~\citep{debruijne99}. The Cepheus OB3 association is one of
the most distant associations currently known, with a distance modulus
of $9.65$ magnitudes (distance $= 851$pc), typical extinction A$_{\rm
V} = 2.81$ magnitudes, an age of less than $10$ Myr and solar
metallicity~\citep{pozzo03}. We use the Padova stellar isochrones
of~\cite{girardi00} assuming solar metallicity ($Z=0.02$) and an age
of $10$Myr to determine the stellar mass corresponding to the faintest
magnitude at which the RVS will obtain velocities to an accuracy of
$3\kms$. For K1III stars the magnitude limit is $V=16$ and for F2II
stars it is $V=14$. In the case of Scorpius OB2, we find that the RVS
will be able to obtain velocities for stars with masses down to
$0.4-0.7$M$_\odot$ depending on the extinction. In the more distant
Cepheus OB3 this mass limit increases to about $2.2$M$_\odot$ due to
both its larger distance and higher levels of extinction. The
availability of velocities for the lower mass stars in associations
will allow more precise determination of the mean motion of the
associations which in turn will permit better discrimination between
members and non-members of all masses.

The RVS data set will dramatically increase the sample of OB
associations which can be studied in detail. There are currently about
$31$ known OB associations within $1.5$ kpc of the Sun and which will
be easily accessible to the RVS. Of these only $12$ have been studied
using a combination of {\it Hipparcos} data and ground-based radial
velocities~\citep{dZHBBB99}.  In fact, $6$ of the associations
identified by~\cite{dZHBBB99} were not included in previous catalogues
which illustrates both the difficulty of identifying associations
without full kinematic information and the likelihood that the number
amenable to study with the RVS will be larger than expected.

\subsection{Open Clusters}

Somewhat more massive than OB associations, with gravitating masses
sufficient to retain a bound remnant following the expulsion of the
gas left over from star formation, the open clusters are important
tracers of the young and intermediate-age stellar populations of the
Galactic disc. Their value derives from the increased accuracy with
which the distances, ages and metallicities of clusters can be
determined compared to those of individual stars. Observations of a
large sample of open clusters will make it possible to look for
metallicity gradients in the Galactic disc which may be preserved more
strongly in the open cluster distribution than in the distribution of
individual stellar
metallicities~\citep[e.g.][]{friel02,Brown01a}. Their internal
velocity dispersions are generally a few $\kms$ although some
(e.g. the Hyades cluster) can be less than
$1\kms$. Recently,~\cite{kharchenko05} have used an all-sky stellar
catalogue to determine the observed properties of a sample of 513 open
clusters (each containing at least 18 members) within about $4$pc of
the Sun. The limiting magnitude in this study was $V\simeq14$ and the
authors estimate that their cluster sample is complete to a distance
of about $1$ kpc.

{\it Gaia} will be able to observe many open clusters -- all known
clusters will be amenable to study and it is expected that several
thousand more will be identified within $5$ kpc of the sun. The
observability of an individual cluster depends on many factors such as
age, distance and extinction, all of which affect both the numbers of
stars with apparent magnitudes brighter than $V=17$, and the velocity
of the cluster relative to the local standard of rest, which can
influence the level of contamination by foreground stars. For objects
whose motion lies mostly along the line of sight, stellar proper
motions will be small and radial velocities will place the tightest
constraints on the membership lists. Thus, as in the case of OB
associations, radial velocities are essential for the construction of
an accurate census of members.

For the Hyades cluster there are almost $400$ stars brighter than
$V=17$ at its present location of $46.34$pc~\citep{perryman98}. For a
similar cluster at a distance of $500$ pc some $200$ stars would still
be observable by the RVS. At this distance the accuracy of the proper
motions will be about $0.5 \kms$ and so the goal of achieving a
comparable radial velocity error with the RVS is achievable for only a
tiny minority of member stars. For these stars, however, it will be
possible to trace their orbits backwards in time and hence estimate a
kinematic age for the cluster by determining their epoch of smallest
separation.

For nearby open clusters the RVS will provide radial velocities with
errors below $1 \kms$ for large numbers of member stars. Even for the
more distant clusters, however, the offset between the radial velocity
of the cluster and the radial velocities of field stars along the line
of sight means that the RVS radial velocities will generally be
valuable in the construction of uncontaminated membership lists. In
addition, stars which have escaped from the cluster but which remain
on similar orbits within the Galaxy to that of the cluster centre of
mass will be identifiable. These stars are vital to our understanding
of the disruption processes which affect open
clusters~\citep[e.g.][]{dlfm96,pzmhm01}. For example,~\cite{dlfm96}
found that in $N$-body simulations of open cluster dissolution the
remnant of an open cluster which is left once the cluster has
evaporated is very rich in binaries. Radial velocities will make it
possible to confirm this result. The long term evolution of clusters
in the Galactic disc is also of great interest as there is evidence
that some of them, for example Cygnus OB2, are as massive as globular
clusters~\citep{kn00}. In this context, the existence of open clusters
such as Berkeley~17 with ages of about $9$ Gyr~\citep{cvgr99} allows
us to investigate the long-term evolution of clusters in a strong
tidal field.

Young open clusters are valuable tracers of recent star formation and
have been used to identify the spiral arms of the Milky
Way~\citep[e.g.][]{feinstein94}. Their eventual dissociation builds up
the population of field stars in the Galaxy and thus knowledge of the
properties of as many such objects as possible is an important step
towards understanding how the stellar populations in the Galaxy formed
and subsequently evolved. As well as representing the dominant mode of
current star formation in the Galactic disc, it has been suggested
that the larger velocity dispersion of the thick disc compared to that
of the thin disc might be partly due to the former containing a
kinematically hot population of stars from clusters which became
unbound by rapid gas expulsion~\citep{kroupa02}. The resultant
characteristic mass function of young clusters constitutes an
important test for this model~\citep{kroupaboily02}. A determination
of the mass function of young clusters is also of fundamental
importance if we want to understand the formation processes of
clusters themselves -- a complete census of cluster members for a
large sample of young clusters is essential for this work. The cluster
mass function also has implications for the chemical evolution of our
Galaxy and others because simulations have shown that long-lived,
massive clusters can significantly enhance the rate of Type Ia
supernovae~\citep{sharahurley02}.

The RVS will be able to test models of cluster disruption by looking
for the unbound populations of stars surrounding young clusters which
are the signature of the effects of gas expulsion. These stars will be
easily identifiable due to the similarity of their metallicities and
space motions with those of their parent cluster. In addition, the RVS
will be able to detect the streams in the thick and thin disc
populations which are expected to result from recently disrupted
clusters. When a cluster becomes unbound, differential rotation and
disc heating mechanisms spread the cluster stars within an elongated
volume centred on the original cluster orbit. The total velocity
dispersion of the stars increases with time -- however, stars which
are physically close together will have a lower velocity dispersion
than the initial value as a consequence of Liouville's
theorem~\citep[e.g.][]{bt87}. As Eq.~\ref{eq:eps_stream} shows, the
velocity accuracy required to distinguish more than a few tens of
streams is difficult to attain for a disc population with the
dispersion of the thin disc. However, for an object which was
disrupted during the past few hundred Myr there will be sufficient
stars moving on orbits similar to that of the original cluster, and
sharing common characteristics such as metallicity, to permit
identification of the remnant.

\subsection{Runaway stars}
Runaway stars are isolated, early type stars (spectral types O and B)
with large peculiar velocities relative to the mean Galactic
rotation. They are thought to originate in associations and the
feasibility of tracing their orbits backwards in time in order to
determine the cluster in which they formed has been demonstrated
by~\cite{hooger00} using a combination of {\it Hipparcos} data and
radio observations. Two mechanisms which lead to the ejection of
runaway stars have been suggested. One possibility is that close
dynamical encounters within an association can result in stars
achieving escape velocity. An alternative channel is the
binary-supernova scenario in which the runaway star was originally a
member of a close binary. Following the explosion of its companion as
a supernova, and possibly the consequent disruption of the binary, the
runaway star acquires a sufficient velocity to move out of the
cluster. Both mechanisms appear to occur in nature, but their relative
importance has not yet been clearly
established~\citep{hooger00}. Retracing the orbits of large numbers of
runaways is essential for the identification of the dominant
production channel. This will, in turn, have implications for our
understanding both of the internal dynamics of associations and
clusters and of the fraction of high mass stars which form in
binaries~\citep{spz00}. It will also improve our knowledge of the
distribution of kick velocities acquired by pulsars at formation.

Radial velocities are useful in this work for two reasons. First,
knowledge of the full-space motion of the runaways makes the
determination of their point of origin significantly more
robust. Secondly, at present the identification of runaway stars
relies on their large proper motions. The inclusion of radial velocity
information will allow their identification based on a true space
motion which may not lie in the plane of the sky -- the RVS will
therefore lead to a significant increase in the detectable sample of
runaway stars.

Another interesting possibility for the post-{\it Gaia} modelling of
the Milky Way will be to use runaway objects as tracers of the
gravitational potential of the disc. An example of such an object is
Cygnus-X2 which~\cite{kdkr00} have suggested may have originated as an
intermediate-mass X-ray binary in the Galactic disc which was
subsequently driven to its present location $2.28$ kpc out of the
Galactic plane by the velocity kick generated in a supernova
explosion. More recently, \cite{brown05} have identified a
hyper-velocity star moving at more than $700\kms$ away from the Galaxy
and currently located at about $55$ kpc from the Galactic centre --
its properties (including its age and metallicity) are consistent with
its having been ejected from the Galactic bulge about $80$ Myr
ago. The {\it Gaia} data set will permit the identification of stars
with anomalously high velocities compared to other stars in their
vicinity. Determination of the orbits of a significant sample of such
stars using both radial velocities and proper motions may be used to
place constraints (albeit somewhat circumstantial ones) on the
distribution of mass in the Galaxy.

\subsection{Globular Clusters}
The globular clusters which orbit the Milky Way are valuable
laboratories in which to study the interplay between stellar evolution
and stellar dynamics. Although the central regions of many Milky Way
globular clusters will not be observable using the RVS due to
crowding, the RVS will nevertheless contribute to our understanding of
these systems in a number of ways. For example, the outer regions of
many clusters will be amenable to study and therefore the internal
dynamics at intermediate radii can be investigated.

It is reasonable to assume that the dense regions of globular clusters
will have a similar impact on the performance of the RVS to that of
crowded Galactic fields of comparable surface brightness. As we
discussed in Section~\ref{sec:perform}, radial velocity estimates
start to be degraded for stellar densities above $2\times 10^4$ stars
per square degree. For field stars, a limiting magnitude of V$=17$
(i.e. a stellar density of about $2\times 10^4$ stars per square
degree brighter than ${\rm V}=17$) corresponds to an integrated
surface brightness (for all stars in the magnitude range V$=10-20$) of
V $= 21.75$ mag arcsec$^{-2}$
\citep[see][]{zwitter490,2003gsst.conf..493Z}. Thus, stars with V$=17$
will be observable in those regions of a globular cluster where the
integrated surface brightness is fainter than 21.75 mag arcsec$^{-2}$.

In order to determine which Galactic clusters will be observable, we
assume further that a star with V$<17$ will be observable against a
given background if the difference between the magnitude of the star
and the integrated surface brightness is the same as that between a
V$=17$ star and the V$=21.75$ mag arcsec$^{-2}$ critical surface
brightness. This implies that, for example, stars in the ranges
V$=14-15$ and V$=15-16$ can be observed against backgrounds of
V$=19.75$ and V$=20.75$ mag arcsec$^{-2}$, respectively. This
assumption of a constant offset is probably conservative, since the
S/N of the source spectrum increases for brighter sources -- it may
therefore be possible to measure velocities for the brighter stars
against higher background densities. From the globular cluster
catalogue of \cite{Harris1996} we extract the concentration and
central surface brightness for each cluster and compute the radii
within which the surface brightness is higher than the RVS limits for
stars in a number of magnitude bins, assuming that the cluster can be
represented by a King profile. We then calculate the proportions of
the total number of stars in the cluster which lie outside these
radii. Finally, using the distance, reddening, luminosity function and
total mass of each cluster, we compute the number of observable stars
in each bin. The cluster masses are estimated from their total
luminosities using a linear interpolation between the 55 clusters for
which mass estimates are available in the literature. The luminosity
function of M92 is assumed for all the clusters. It is important to
note that we have ignored the fact that for some RVS transits,
individual spectra may overlap denser regions of the cluster, thereby
reducing the effective number of transits per star. In view of this
and our other simplifying assumptions, the estimates are probably only
reliable to within about a factor 2-3. Nevertheless they give an
indication of the likely contribution of the RVS to globular cluster
studies.

\begin{figure}
\includegraphics[width=0.5\textwidth]{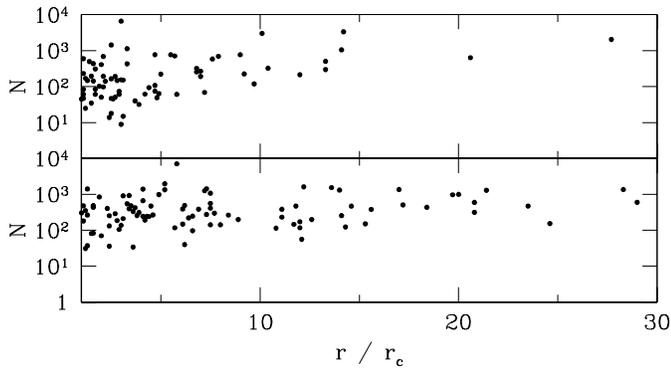}
\caption{Performance of the RVS in Galactic globular clusters. For
each cluster, the plot shows the innermost radius $r$ (in units of the
cluster core radius $r_{\rm c}$) at which observations will be
possible for stars of V$=15$ (98 clusters: top panel) and V$=17$ (121
clusters: bottom panel) and the total number of stars $N$ which will
be observed in the cluster with V$<15$ (top panel) and V$=15-17$
(bottom panel). Clusters which either cannot be observed or in which
the innermost radius is more than 30 core radii have been omitted. See
text for a detailed discussion.}
\label{fig:GC_results}
\end{figure}
Fig.~\ref{fig:GC_results} summarises the performance of the RVS in the
Milky Way globular cluster population. In approximately 33 clusters,
the RVS will be able to observe more than one hundred stars brighter
than V=15 to within 5 core radii -- for 24 clusters it will be
possible to observe stars inside one core radius. The individual
radial velocity errors for these stars will, in most cases, be smaller
than the internal dispersion of the cluster, making them extremely
useful for the discussion of the internal dynamics of the
cluster. Thus the RVS will provide a large sample of stars for the
detailed study of the internal kinematics of about 20-30 of the
Galactic globular clusters. For 96 clusters, more than one hundred
stars in the range V=15-17 will be observable to radii of less than 5
core radii. These latter stars will be useful for determining the bulk
properties of the clusters such as rotation.

One key quantity which the RVS will constrain is the stellar binary
fraction in globular clusters which, despite its importance for
understanding the dynamical evolution of
clusters~\citep[e.g.][]{hut92,wilk03}, is currently poorly constrained
observationally~\citep[e.g][]{elson98,albrow01,stetson03}. While we
expect that the fraction of tight stellar binaries should decrease
with increasing radius in most clusters due to the effects of mass
segregation, a limit on the binary fraction in the outer parts will at
least provide a constraint on the fraction in the cluster as a
whole. Cluster binaries will also be detected using the astrometric
and photometric data from \textit{Gaia}.

Accurate velocities for stars in the outer parts of clusters will
permit the study of rotation in clusters as well as providing
estimates for the bulk radial velocities of the clusters. In
conjunction with the {\it Gaia} proper motions for the clusters, these
will determine the orbits of all the Milky Way clusters and place
tighter constraints on the mass of the dark halo~\citep[e.g.][ and
\S~\ref{sec:dark_matter}]{we99}.

The gravitational field of the Milky Way continually removes stars
from the outer parts of the star clusters which orbit the Galaxy,
leading to the development of tidal ``tails'' of stripped stars which
extend along the orbits of the clusters. There is some evidence for the
presence of tails around a number of globular
clusters~\citep[e.g.][]{lmc00,odenk01}. However, recent work has
called into question the reality of certain of these tidal tails and,
in particular, suggests that the tails of the massive globular cluster
Omega Centauri are artifacts produced by differential reddening across
the face of the cluster~\citep{law03}. Kinematic studies of the outer
parts of clusters are necessary in order to improve our understanding
of cluster disruption by external tides.

The RVS spectra will contribute to the resolution of this issue in two
ways. First, as we discuss in \S~\ref{sec:extinction_maps}, the
spectra will facilitate the construction of accurate extinction maps
throughout the Galaxy. Secondly, the velocities of stars in the tidal
tails of a star cluster are similar to the bulk motion of the cluster
and, in general, differ significantly from those of surrounding field
stars. As a result, the velocity estimates provided by the RVS (and,
of course, by the {\it Gaia} proper motions) will often be sufficient
to distinguish between escaping cluster members and field stars.
Thirdly, the metallicities of cluster stars will generally differ from
those of the surrounding field stars. For brighter stars, this will
provide another potential confirmation of a physical connection, if
any exists, between the ``tails'' and the cluster. 

All Galactic clusters with sufficient numbers of bright (V$<17.5$)
stars will be amenable to study via a combination of the above
approaches; for example, the well-studied cluster Pal~5, whose tidal
tails have now been traced in the stellar number density distribution
to distances of about $2$ kpc from the cluster
centre~\citep{odenk03}. While the internal velocity dispersion of the
tails is too small for the RVS velocities to contribute to discussions
of their internal structure (the intrinsic dispersion of Pal~5 is less
than $1\kms$~\citep{odenk02} and the spatial coherence of the streams
strongly suggests that they must also be kinematically cold), the
velocities will be of sufficient accuracy to distinguish tail members
from Galactic field stars. This will be of particular value for those
clusters whose tidal tails are less extended or less clearly defined
spatially than those of Pal~5. It is also worth noting that {\it Gaia}
will also contribute to the identification of tidal streams by means
of proper motions, since stars in tidal tails display proper motions
very similar to those of cluster members and, in general, quite
different from the field stars along the line of
sight~\citep[e.g.][]{king98}.

\subsection{Chemical Evolution}

To build a realistic model of the chemical evolution of the Galaxy one
needs as many observational constraints as
possible~\citep{pagel97}. These observations not only have to include
the relative numbers of stars in the thin and thick disc or the halo,
but also have to be able, using the stellar abundances of some
chemical elements, to constrain particular model inputs, for example
the rates of type I and II supernovae. Possible by-products of these
chemical abundance measurements could include the determination of an
Age-Metallicity relation and possible gradients in the thin and thick
discs (see also \S~\ref{sec:AMR}). The RVS spectra will reveal mainly
[$\alpha$ element/Fe] ratios for late type stars (e.g. the $\alpha$
elements Ca, Mg, Si, Ti), which constrain the sites of nucleosynthesis
-- these stars will require follow-up observations, for example, of
the r process elements not obtainable from the RVS spectra.  The RVS
will provide precise chemical abundances for $3-6\times10^6$ stars
brighter than V=12-13. This will make it possible to probe the
thin/thick disc and halo populations up to a distance of 2 kpc when
K1III stars having absolute magnitude M$_{\rm V} \sim 1$ are used as
tracers and up to 300 pc using G2V stars of M$_{\rm V}=5$.

The variation of [$\alpha$/Fe] among stars as a function of their
location in the Galaxy will most strongly constrain models of the
quantity of gas in the Galaxy at different epochs and also the
supernova rate. The combination of chemical abundances and kinematical
properties has revealed some difficulties with naive infall models of
the Galaxy as was first demonstrated by~\cite{nissen97}. It now seems
difficult to support a simple infall model: this adds weight to the
introduction of a more violent scenario of the formation of the
Galaxy~\citep[e.g][]{hw99}. However, in such a scenario it is not
obvious that one would expect a well-defined relation between chemical
abundance ratios and age~\citep{ryan03} during the epoch when Type Ia
supernovae are supposed to explode ($\rm [Fe/H]\sim -1.0$).

Searching for extreme Population~II stars is also a challenge for the
{\it Gaia} survey in order to determine the properties of the most
metal-poor stars. These fossil stars are representative of the first
phase of the Galaxy just after the hypothetical era of the
Population~III stars. Observing their [C/O] ratios with ground-based
spectroscopy makes it possible to look for evidence of the presence of
Population~III stars very early on in the history of the
Galaxy. \cite{akerman04} have discovered a possible increase of the
[C/O] ratios for very metal poor stars which would require the
presence of massive Population~III stars at the beginning of the
formation of the Galaxy.  Calcium lines will be visible even in the
RVS spectra of very metal poor stars. The Ca line at $\lambda=854.2$
nm can be seen with an equivalent width EW $= 0.0482$ nm at
[Ca/H]$=-3$ and EW $= 0.0245$ nm at [Ca/H]$=-4$ for a giant star with
$T_{\rm eff}=4800$ K, $\log g=1.5$. This should allow the separation of a
giant with [Ca/H]$=-4$ from a giant of [Ca/H]$=-3$ down to about a
S/N$\sim10$ corresponding to V$\sim15$ for the above types of
star. Assuming this star is a giant of absolute magnitude M$_{\rm
V}=0-1$ then a survey for stars more metal poor than [Ca/H]$=-3.5$
will be possible within a sphere of 5-10 kpc.  {\it Gaia} will also be
able to detect binarity among such halo stars and hence rule out mass
transfer scenarios through which the carbon abundance in the
primordial atmosphere of an extreme-Population~II star is modified,
thereby producing an incorrect diagnostic on the presence of
Population~III stars in the early phase~\citep[see
e.g.][]{2003sf2a.conf..547M}.

Among the successes of the standard big-bang model and of the
\textit{Wilkinson Microwave Anisotropy Probe} (WMAP)
experiment~\citep{spergel03} are the stringent constraints on
predictions of primordial light element abundances. In particular, the
helium abundance $\rm Y$ is now constrained to lie in the range
$0.247-0.252$. This value is then supposed to be the primordial helium
content of the Galaxy. Its enrichment with time is assumed to be a
consequence of multiple generations of stars and is expected to follow
the enrichment of heavy elements Z. However, the details of the helium
to metal enrichment relation $\delta Y / \delta Z$ which describes the
increase in helium from primordial levels have long been debated in
the literature~\citep[see][ for a review]{hog98}. {\it Gaia} will
provide a survey of several thousand binaries many of which will have
accurate age estimates because when the masses of the components of a
binary are known with accurate effective temperature, luminosity and
abundance Z then the age can be determined by fitting evolutionary
tracks to the two error boxes of the binary in the HR diagram.  The
ages and estimates of He content deduced from these Hr diagram fits,
can be used to discuss the helium enrichment of individual systems:
for example, \cite{lebreton01} present an analysis of five binaries in
the Hyades cluster and determine the helium content and age of the
cluster. The Galactic enrichment relation $\delta Y / \delta Z$ will
be constrained not only for close binaries with Solar metallicities
but also for very metal-poor binary stars.

At the same time, {\it Gaia} will provide the opportunity to measure
magnitudes, absolute distances and chemical compositions for a
substantial fraction of halo and thick disc low-mass Horizontal Branch
(HB) and RGB stars. These new data will supply accurate estimates of
the R parameter, i.e. the ratio between the number of HB and RGB
stars~\citep{iben1968}. The comparison between empirical star counts
and evolutionary lifetimes provides an estimate of the Helium content
and in turn an upper limit on the primordial Helium
abundance~\citep{zoccali00}. Up to now the R parameter has been
estimated only in globular clusters and current estimates are
partially hampered by statistics and by the accuracy of spectroscopic
measurements~\citep{cassisi03}. The new Helium abundances will also
provide the opportunity to estimate $\delta Y / \delta Z$ in the
Galactic halo and thick disc as a function of Galactocentric distance.
 
\subsection{Extinction maps}
\label{sec:extinction_maps}
\begin{figure}
\includegraphics[width=0.5\textwidth]{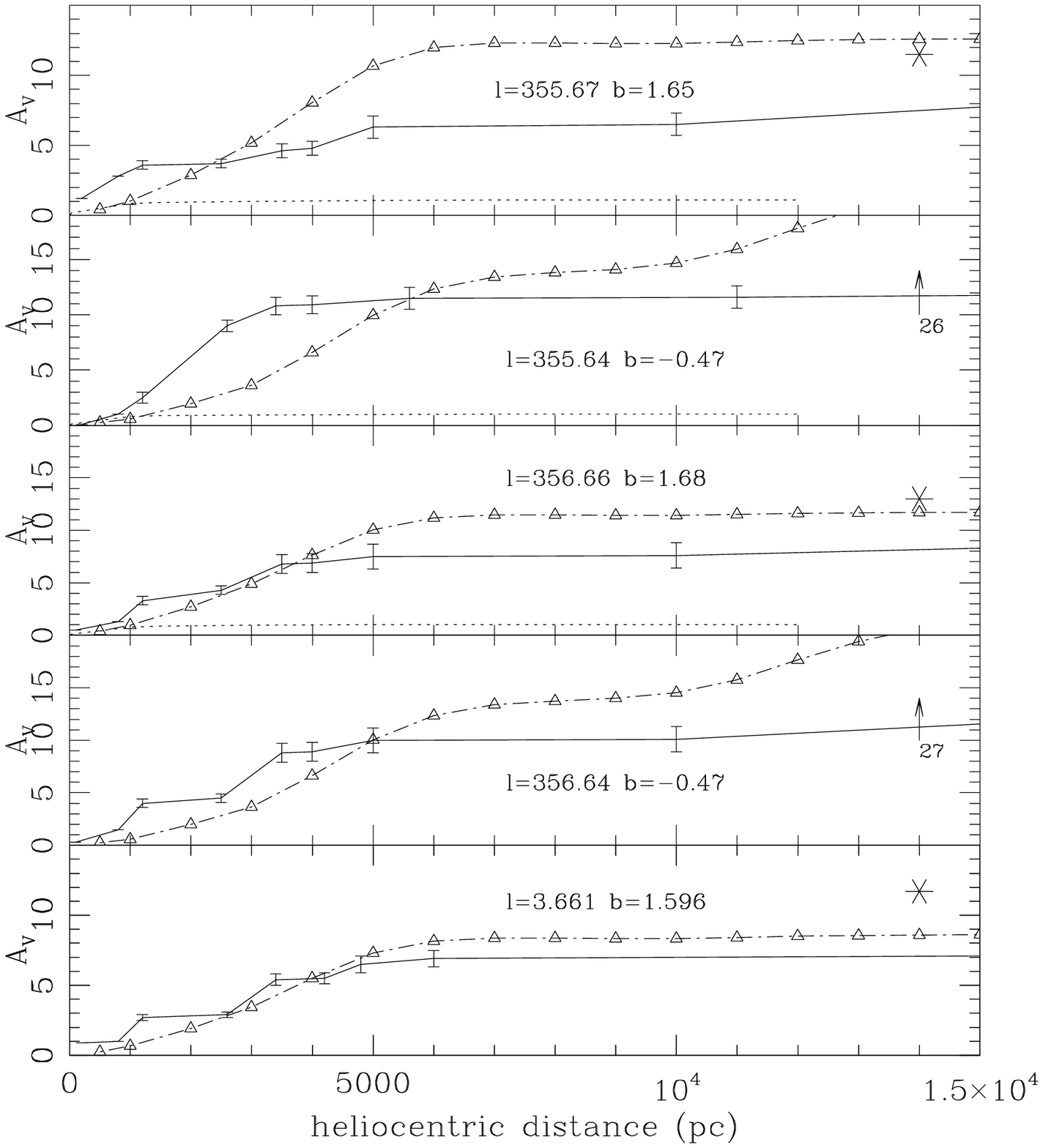}
\caption{Extinction along several lines of sight towards the Galactic
centre. The~\protect\cite{vallenari03} determination (solid line in
each panel) is the mean value in a field of
$0.5^\circ\times0.5^\circ$.  The error bars indicate variable
reddening inside the field. The dotted line represents the model
of~\protect\cite{1998A&A...330..910M}; the dot-dashed line with
triangles is the~\protect\cite{2001ApJ...556..181D} model including a
rescaling factor. The star gives the value from
the~\protect\cite{1998ApJ...500..525S} maps. When the determination
from the~\protect\citeauthor{1998ApJ...500..525S} maps lies outside
the plotted range, a labelled arrow indicates the reddening value.}
\label{maps}
\end{figure}

Interstellar absorption represents one of the major complications in
the simulation of the CMDs of Galactic stellar populations along any
line of sight, especially at low Galactic latitudes across the
disc. The uncertainties in star count predictions caused by reddening
may amount to some $26$ per cent \citep{1999A&A...352..459C}.

Several empirical models of Galactic absorption are available in the
literature.  \cite{1998A&A...330..910M} make use of the large-scale
properties of the dust layer in the Galaxy to derive the absorption in
the Galactic plane. \cite{1998ApJ...500..525S} use the
\textit{COBE}/DIRBE 100 $\mu$m and 240 $\mu$m data to construct maps
of the dust temperature in the Galaxy.  In high-latitude regions, the
dust map correlates well with maps of H~I emission, but deviations are
found in parts of the sky and are especially conspicuous in regions of
saturated H~I emission towards denser clouds and in areas of H$_2$
formation in molecular clouds.  Recently, \cite{2001ApJ...556..181D}
and~\cite{2003A&A...409..205D} presented a three-dimensional model of
the dust distribution based on COBE/DIRBE infrared data.  As stated by
~\cite{2003A&A...409..205D}, regions having anomalous emission due to
warm dust are not well described by the model.

When comparing the various literature models of extinction in analyses
of CMDs observed along the line of sight, very large discrepancies
soon become apparent.  First, there are still problems with the
accuracy of the zero point calibration of the extinction maps.  In
fact, as \cite{burstein03} points out, even at high Galactic latitudes
the reddening map by~\cite{bursteinheiles78} provides reddening values
that are 0.02 mag smaller than the reddening map
by~\cite{1998ApJ...500..525S}.  At low Galactic latitudes and towards
the spiral arms, the difference may be as large as 4-5
magnitudes~\citep[see Fig.~\ref{maps} and][]{vallenari03}. This is
due in part to variations of the dust properties on small scales and
in part to the many uncertainties which are still associated with dust
emission models.

\begin{figure}
\begin{center}
\includegraphics[width=0.4\textwidth]{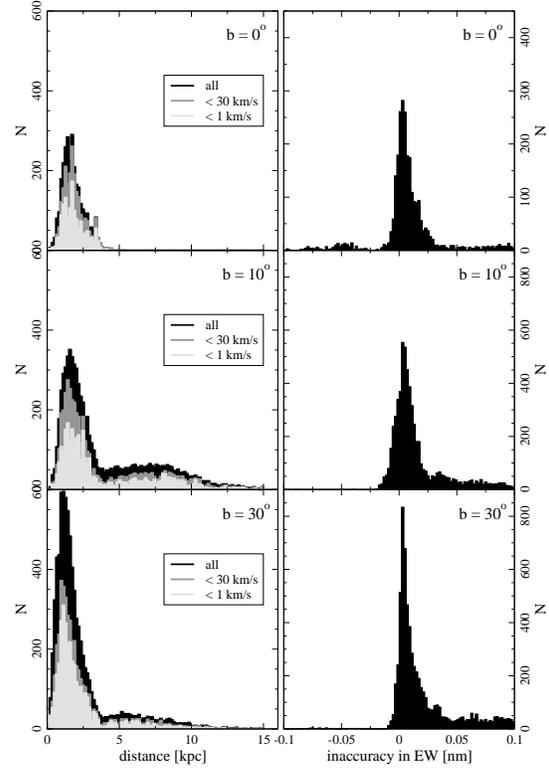}
\caption{\label{dib_figure} Recovery of the $862$ nm DIB from
simulated RVS spectra. The stellar spectra are obtained from the
stellar population synthesis Galactic model
of~\protect\cite{rrdp03}. The three-dimensional model of the dust
distribution of~\protect\cite{2001ApJ...556..181D} is
used. \textbf{Left:} Number of simulated stars as a function of
distance in three Galactic directions ($l=10^\circ$; $b=0^\circ$,
$b=10^\circ$ and $b=30^\circ$). Black represents all simulated stars,
dark grey those for which the DIB line was recovered with an accuracy
of at least $30\kms$ and light grey those for which the recovery of
the DIB line was more accurate than $1\kms$. As expected, extinction
is a strong limiting factor only within the Galactic
plane. \textbf{Right:} Number of simulated stars as a function of
inaccuracy in equivalent width EW, i.e. the difference between the
simulated and recovered EW of the DIB line. According
to~\protect\cite{munariDIB00}, EW is proportional to $E_{\rm B-V}$
(an EW of 0.05 nm corresponds to a reddening $E_{\rm B-V}=1.35$).}
\end{center}
\end{figure}

The RVS can directly measure an interstellar or circumstellar
extinction corresponding to $E_{B-V}=0.10$ using the $862$ nm diffuse
interstellar band (DIB: see Paper~I) of early type stars brighter than
$V\sim12-13$ which has been shown to be directly associated with the
dust phase of the interstellar medium.  Simulated RVS data indicate
(see Fig.~\ref{dib_figure}) that the imprint of the $862$ nm DIB will
be detected even in the spectra of much fainter stars with magnitudes
up to $V\sim 16$ with sufficient accuracy to trace not only the
distribution of the interstellar medium but also the radial component
of its kinematic motion (the Doppler velocity of the mass center of
the dust cloud is calculated from the wavelength position of the DIB
line centre). The resulting three-dimensional maps of interstellar
extinction are independent of the photometric approach to reddening
determination which is based on a comparison between observed colours
and modelled intrinsic colours. Thus, the RVS will make a vital
contribution to the construction of a new generation of accurate star
count maps of the Galaxy which are essential to the development of
complete models of Galactic structure. In addition, accurate
extinction-corrected stellar magnitudes will be of enormous importance
for the testing of stellar models (see
\S~\ref{sec:stellar_evolution}), solving the degeneracy between
reddening and temperature that usually affects determinations from
photometry.

\begin{table*}
\begin{center}
\begin{tabular}{lccccccc}
Field & FOV (arcmin$^2$) & RA(J2000) & DEC(J2000) & l & b & A$_{\rm B}$ & Ns \\ \hline
L95   &  $46.0\times50.9$   &  03 54 12.40  & +00 13 09.0   &  188.79146  &  -38.25715  & 0.67/1.51  & 870/1090\\
L107  &  $44.1\times37.5$  &   15 39 20.90  & -00 20 25.9    &   5.70550   &  41.22154  & 0.29/0.48 & 1170/1525\\ 
MarkA  & $33.9\times26.9$   &  20 43 43.20  & -10 46 16.9   &   35.89721   & -29.77091  & 0.18/0.25 & 1700/2215\\ \hline
\end{tabular}
\end{center}
\caption{Parameters of selected Galactic fields which illustrate the
range of conditions which the RVS will encounter at intermediate
latitudes. The table presents the field name, the field of view for
each observed field in square arcmin, the positions of the field
centres in ecliptic and Galactic coordinates, the interstellar
extinctions A$_{\rm B}$ estimated over an area of 5$\times$5 arcmin
centred on the individual fields and the estimated number of stars in
each field $N_{\rm s}$ which will be detectable by the RVS. Note: Two
values of A$_{\rm B}$ (and the associated $N_{\rm s}$ estimates) are
quoted for each field. The $A_{\rm B}$ values are taken from
the~\protect\cite{burstein82} and~\protect\cite{1998ApJ...500..525S}
reddening maps, respectively.}
\label{tab:disc_fields}
\end{table*}

As an illustration of the effects of reddening on the stellar
distribution in different regions of the Galactic spheroid we selected
a few well-observed fields.  In particular, we focused our attention
on three standard stellar fields for which~\cite{stetson00} (see also
{\tt http://cadcwww.hia.nrc.ca/standards/}) collected new and
homogeneous multi-band photometric data. Table~\ref{tab:disc_fields}
lists the coordinates and positional parameters of the selected
fields.

\begin{figure}
\includegraphics[width=0.5\textwidth]{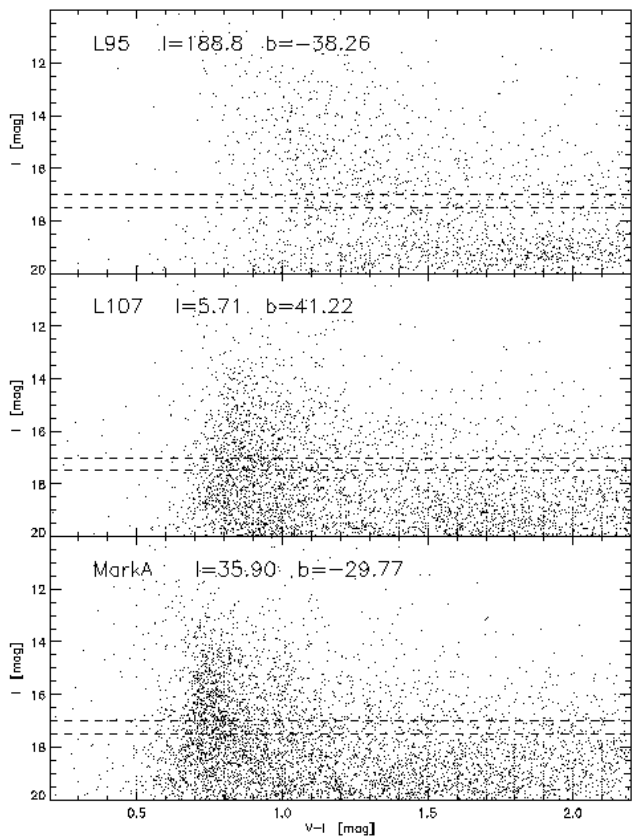}
\caption{($V-I,I$) CMDs for the three fields listed in
Table~\ref{tab:disc_fields}. The dotted horizontal lines show the
likely range of the faint magnitude cut-off for the RVS radial
velocity determinations.}
\label{fig:disc_fields}
\end{figure}

Fig.~\ref{fig:disc_fields} shows the ($V-I,I$) CMDs of the selected
fields which are assumed to be representative of the number densities
and extinctions for intermediate latitude Galactic fields. The data
plotted in this Figure clearly show the effects of reddening.  When
moving from bottom to top, the reddening increases from E(B-V)=0.04 to
0.16: the increase, as expected, causes a systematic shift towards
fainter magnitudes and redder colours for the field stars. This means
that observed luminosity functions and colour distributions are
strongly affected by the interstellar extinction and by differential
extinction, if any, along the line of sight. We note in passing the
large difference in the mean interstellar extinction between
the~\cite{burstein82} and the~\cite{1998ApJ...500..525S} reddening
maps (see column 7 in Table~\ref{tab:disc_fields}). This emphasises
again the crucial role of individual reddening measurements in the
improvement of the accuracy of stellar parameters.

\subsection{Star formation history}
The determination of the history of star formation from the CMDs of
composite stellar populations is an important goal of modern
astrophysics.  The problem is easier to tackle in galaxies in which
individual stars are resolved and CMDs are derived, because we may
assume that all the stars lie at almost the same distance. However, in
our own Galaxy the problem is significantly more complicated because
there are differences in the distances of the stars and therefore only
CMDs containing stars of inhomogeneous age, chemical composition and
distance are directly exploitable. Additionally, the age-metallicity
degeneracy further complicates the problem.  Following the {\it
Hipparcos} mission, it was possible for the first time to derive the
CMD, in absolute magnitudes, of field stars in the Solar vicinity
\citep{1995A&A...304...69P} and from this CMD to study the history of
the Solar neighbourhood.

The {\it Gaia} mission will significantly exceed the performances of
{\it Hipparcos} permitting the determination of the star formation
histories (SFH) of the disc and halo of the Milky Way.  Recently,
\cite{2001AJ....121.1013B} determined the SFH of the Solar vicinity
from {\it Hipparcos} data, covering the total lifetime of the disc (10
Gyr).  They find that from the {\it Hipparcos} catalogue it is
possible to select a complete sample of stars with well measured
parallaxes down to M$_{\rm V}=4.5$ in order to include main-sequence
and evolved stars, inside a sphere of radius $r=50\,$pc.  {\it Gaia}
will be able to observe a much larger sample of stars covering in
distance and position a large portion of the Galaxy and possessing the
same degree of accuracy as that obtained by {\it
Hipparcos}. \cite{2002EAS.....2..265B} has demonstrated that {\it
Gaia} will indeed allow the study of a sample of stars complete down
to M$_{\rm V}=4.5$ with a parallax accuracy better than
$\sigma_\pi/\pi\leq0.1$ up to a distance of $2-3$ kpc. Here it is
worth remembering that due to the effects of radial mixing in the disc
as a response to spiral wave perturbations, old stars born at the
location of the Sun may now be spread over a range of Galactocentric
radii from 4 to 12 kpc~\citep{sellbinney}.  Within this volume, the
RVS will provide information about the average metal content of the
stars brighter than about $V = 14-15$ (see Table~\ref{tab:performance}
and Paper~I), solving the age-metallicity degeneracy which always
hampers the estimation of the star formation rate from
Hertzsprung-Russell (HR) diagrams, in particular for older stars.
Additionally, information on the metal content of fainter stars will
be provided by {\it Gaia} photometry and astrometry.  Since different
populations have different kinematics and metallicities, coupling
these properties would allow us to distinguish statistically between
thin/thick disc and halo stars.
 
\subsection{Age-metallicity relation}
\label{sec:AMR}
The age-metallicity relation (AMR) for disc stars, if such a relation
exists, gives information about the process of stellar formation,
stellar orbit diffusion from scattering by molecular clouds
\citep{1993A&A...280..136F,1993A&A...275..101E} and about the time
scale for gas mixing \citep{1997A&A...318..231V}.  The existence of an
AMR has long been a controversial issue. \cite{1980ApJ...242..242T}
found an AMR in field stars, while \cite{1993A&A...275..101E} derive
no AMR from a sample of about 187 FG giants with known metallicity,
distance and magnitude.  \cite{1998A&A...329..943N}, using revised age
estimates, derive a moderate AMR with a slope 0.07 dex/Gyr. These
authors find that the main source of uncertainty in the
age-metallicity determination is due to the distance estimates which
are used in the conversion from apparent to absolute magnitudes. On
the basis of their simulations, distances need to be known with at
least $5$ per cent accuracy to obtain ages with a precision of about
$16$ per cent. \textit{Gaia} will observe stars brighter than M$_{\rm
V}=5$ with a distance accuracy of less than $1$ per cent up to $1$
kpc, and with an accuracy of $5$ per cent up to $2$ kpc.  The results
of \cite{1998A&A...329..943N} are in substantial agreement with the
\cite{2000A&A...358..850R} study of a sample of 552 stars. They derive
metal content information from Str\"{o}mgren photometry and ages from
chromospheric activity, finding an AMR of 0.05 dex/ Gyr.

Based on 5800 stars from the {\it Hipparcos} catalogue with ages
derived from isochrones and metallicities estimated from Str\"{o}mgren
photometry, \cite{2001A&A...377..911F} found an AMR only for objects
younger than 2 Gyr.  Recently, using isochrone fitting to derive ages
and Str\"{o}mgren photometry to estimate the metallicities,
\cite{2004A&A...418..989N} find no AMR on a large sample of 14000 F
and G dwarfs in the solar vicinity.  All the above studies, however,
suffer either from the small sample of well-measured stars
\citep{1993A&A...275..101E,1998A&A...329..943N} or from uncertainties
in the determination of the metal content from Str\"{o}mgren
photometry which is difficult to calibrate~\citep{haywood2002}.  {\it
Gaia} will allow age estimations via accurate positions in the HR
diagram. The determination of the metal content for a large sample of
disc stars which will be possible using the spectra from the RVS will
be an essential complement to these accurate ages in improving our
knowledge of the age-metallicity relation.  

Finally, chromospheric activity can be calibrated as an age indicator
against isochrone fitting. Chromospheric activity is expected to
decline while stars are aging and can therefore be used to date stars,
mainly F, G, and K dwarfs. At present, this method is poorly
calibrated and many uncertainties remain: first, there is intrinsic
variability of stellar activity (e.g. the activity cycle in the Sun);
secondly, stellar activity is caused mainly by rotation which,
although it generally decreases with age, can be influenced by, for
example, tidal interaction in binary systems~\citep{kawaler1989}. The
RVS data set will include thousands of eclipsing binaries (for which
the uncertainty in the projection of the rotation vector along the
line of sight is removed) containing F,G and K dwarfs locked in
synchronous rotation with their orbital motion and spanning a wide
range of orbital periods. It will thus be possible to calibrate
accurately the correlation between their levels of chromospheric
activity (as traced by photometry and CaII emission cores in RVS
spectra) and their rotation speed (see \S~\ref{sec:rotation}). Using
the~\cite{kawaler1989} calibration of rotational velocity against age
for main-sequence stars older than 100 Myr, an uncertainty of $5\kms$
in the rotational velocity will result in an uncertainty of $20$ per
cent in the age of a star with a rotation velocity of $50\kms$.

\subsection{The Galactic Bulge}
\label{sec:bulge}
The derivation of the ages, age distribution, metallicities and
kinematics of the stellar populations of the bulge is extremely
important for our understanding of the process of galaxy formation.
As was discussed earlier, the high stellar densities along lines of
sight which pass within $5-10$ degrees of the Galactic centre render
observations of the bulge using the RVS very difficult. However, the
outer bulge will still be amenable to study by the RVS and much will
be learned about the nature of the bulge/bar from the RVS
observations.

\begin{figure}
\includegraphics[width=0.47\textwidth]{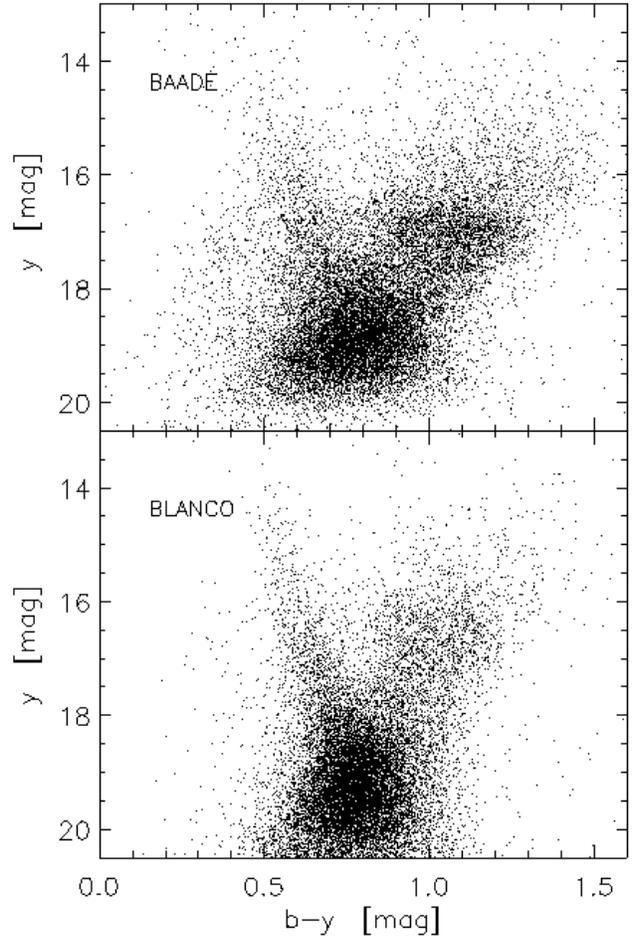}
\caption{CMDs for a field in Baade's window towards the globular
cluster NGC~6522 (top panel) and a region in the Blanco field (bottom
panel).}
\label{fig:bulge_fields}
\end{figure}

To obtain a more quantitative estimate of the number of stars
appearing in the RVS band towards the central regions of the Galactic
bulge we observed two regions, one located at $l=0.99, b=-3.94$ (in
Baade's window) which contains the Globular Cluster NGC~6522 and one
at $l=0.27, b=-6.19$ (the Blanco field) which is one of few regions of
the Galactic bulge that is characterised by low reddening. We
collected a series of CCD images in the Str\"{o}mgren bands {\em
u,v,b,y,Ca} with the Danish 1.5m telescope at ESO La Silla (field of
view $13.7\times13.7$ arcmin$^2$, $0.46$ arcsec/pixel).
Fig.~\ref{fig:bulge_fields} shows the ($b-y,y$) CMD for these
fields. In the Blanco field we detected 2700 stars brighter than ${\rm
y}=17.5$, while the Baade's Window field contains 8900 such stars (we
note that the y band is the closest Str\"{o}mgren filter to the RVS
band); these correspond to densities of about 52\,000 and 170\,000
stars per square degree, respectively. The mean reddening towards both
these fields is E(B-V)$\approx0.45$. The difference in the star counts
is due to the presence in the latter field of NGC~6522 and also to the
fact that in this region we are looking towards the Galactic center.

As the Figure shows, the majority of bulge stars lie well below the
likely magnitude limits of the RVS in these regions. Investigations
are underway to determine strategies for obtaining at least some
information on the brighter stars in certain bulge fields. In the
event that it is possible to obtain these data,
Fig.~\ref{fig:bulge_fields} shows that bulge AGB stars would be
amenable to study at magnitudes between 14 and 15. Given that the line
of sight depth of the bulge is approximately one magnitude, a limit of
14.5 would permit the observation of AGB stars all through the bulge
on these lines of sight.

Even in the absence of data close to the Galactic centre, the RVS will
illuminate certain aspects of the Galactic bulge, in particular the
presence and nature of the Galactic bar. The presence of a bar may be
the origin of some of the stellar moving groups observed in the Solar
neighbourhood. Further, there is growing evidence for the presence of
bulge/bar features in the stellar distributions at larger angular
distances from the Galactic centre. For example,~\cite{hammersley00}
identified a red clump feature at $l = 27^\circ$ and a similar one has
been identified at $l = 9^\circ$ by~\cite{babu05}. This feature, which
is also seen in observations of OH/IR stars~\citep{sevenster99} and
SiO masers~\citep{izumiura99}, may be associated with a second bar,
such as are often observed in external
galaxies~\citep[e.g.][]{erwin03}, or a stellar ring at a distance of
$3$ to $4$ kpc from the Galactic centre.  Along the same lines is the
work by~\cite{2003A&A...408..141P} who detect a star-count excess in
NIR data with respect to the expected disc population at $15^\circ < l
< 27^\circ$.  The radial velocities provided by the RVS will help to
distinguish between the possible explanations of these features. Thus,
although the RVS will at best observe a limited number of bright stars
in the most central regions of the Galaxy (and at worst not observe
the central regions at all), it will nevertheless provide very useful
constraints on the properties of the bar and bulge by observing bulge
features outside $10^\circ$ from the Galactic centre.

\subsection{Local Group Galaxies}
\subsubsection{The Magellanic Clouds}
The Magellanic Clouds provide a unique opportunity to study the
evolution of dwarf irregular galaxies at close quarters as well as the
effects of galaxy interactions (both the interactions between the
clouds themselves and the external effect of the Milky
Way). \cite{spite02} has highlighted a number of outstanding issues
regarding the Magellanic Clouds which can be addressed by the {\it
Gaia} mission. The resolution of one of these in particular relies on
the spectroscopic capabilities of the RVS: does the Large Magellanic
Cloud (LMC) possess a pressure-supported stellar halo? Evidence for
such a population has recently been presented by~\cite{minniti03}
using observations of 43 RR Lyrae stars in the inner regions of the
LMC. The observed velocity dispersion of this stellar halo is $50
\pm10$km\,s$^{-1}$~\citep{minniti03}, which is considerably larger
than the LMC thick disc velocity dispersion of about $20$km\,s$^{-1}$,
estimated from the motions of the old LMC star clusters and
intermediate-age Carbon stars~\citep{vdm02}. The RVS will therefore be
able to distinguish kinematically between members of the two
populations for the RR-Lyrae and metal-poor K giants with magnitudes
in the range 16-17.5 as it will yield velocities for these stars
accurate to better than $15$km\,s$^{-1}$. It will observe stars over
the entire area of the LMC thereby making it possible to determine
whether the population is rotating -- this is not currently
constrained by the data of~\cite{minniti03}. The presence of such a
halo has direct implications for our understanding of the formation
history of the Magellanic Clouds, as it implies a similar hierarchical
evolution to that of the Milky Way. It also has implications for the
interpretation of the microlensing data towards the clouds. For
example, ~\cite{alves04} suggests that two of the known MACHO
microlensing events might be due to lensing of LMC disc stars by stars
in the putative halo, although he points out that the halo cannot
account for all the observed microlensing optical depth. In addition,
a comparison of the kinematics of other stellar populations with those
of the Carbon stars may yield further insights into the nature of the
stellar bar in the LMC, which is currently
unclear~\citep[e.g.][]{ze00}.

The Magellanic Clouds contain a large population of star
clusters~\citep{mg03b,mg03a} with integrated $V$ magnitudes of about
$10-13$. As in the case of the Milky Way globular clusters, these are
useful tracers of the mass distribution and internal kinematics of the
Magellanic clouds. The RVS will be able to obtain accurate radial
velocities for all the clusters yielding a data set of more than $50$
tracers per galaxy. Given that the cluster populations extend to
larger radii than other tracers, these will be invaluable in the
determination of the total gravitating mass of the clouds thereby
constraining models of their formation. The internal dynamics of the
Clouds and their relationship to each other, as well as the role
played by the Milky Way in their evolution, are issues which remain to
be comprehensively addressed.

\subsubsection{The Andromeda Galaxy}
The Andromeda galaxy is the other massive galaxy in the Local
Group. Its stellar halo is currently the focus of considerable
interest due to the presence of large amounts of substructure, in
particular a stellar stream~\citep{ibata_stream01}. Recent estimates
of the total mass of the Andromeda galaxy have shown that there is no
kinematic evidence for the generally held belief that Andromeda is the
most massive galaxy in the Local Group: the data on tracers outside
about $20$ kpc favour a halo which falls off more rapidly outside $30$
kpc than that of the Milky Way~\citep[][]{ew00}. The key difficulty
facing attempts to verify this conclusion is the paucity of tracer
objects in the crucial radius range from $30$ to $100$ kpc. At present
there are only a handful of globular clusters and dwarf galaxies at
these radii. Recent observations of the globular cluster population of
M31 within $25$ kpc have shown that roughly half of the metal-poor
clusters are brighter than $V = 17$~\citep{perrett02}.  Given that
there are more than $400$ confirmed globular clusters within a radius
of about $30$ kpc~\citep{bhbfsg00} compared with about $150$ at all
radii within the Milky Way, it is likely that there will be about $50$
clusters orbiting M31 at radii useful for probing the halo. These
clusters will be spread over a large area of the sky -- spectroscopic
confirmation of the nature of potential cluster candidates is
essential, making this program very time-consuming from the
ground. The accuracy which the RVS will be able to achieve for
integrated spectra has yet to be established. However, velocity errors
of about $10-15\kms$ would be acceptable for constraining the halo
mass of M31. RVS spectra can also be used to identify other likely
cluster candidates which can subsequently be followed-up from the
ground.

Bright ($I \sim 16$) AGB stars belonging to the halo and
intermediate-age disc populations will provide additional tracers of
the M31 potential. The $\sim 10^4$ young disc stars brighter than $I =
16$ and located throughout the M31 disc will permit a detailed
comparison of the stellar kinematics with the well-studied gas
kinematics~\citep[e.g][]{braun91}. They will also facilitate the
detailed modelling of disc features such as the warp.

\subsection{Serendipitous discoveries}
Given the richness of the {\it Gaia} data set, it is to be expected
that it will also throw up many surprises in the field of galactic
dynamics. In addition to revolutionising our existing picture of
dynamical structures such as open clusters, moving groups, etc., many
new features will undoubtedly be found. The scope for such discoveries
is highlighted by a number of recent discoveries relating to the Milky
Way disc, namely (i) the detection of a population of thick disc stars
whose surprising kinematics suggest that they may have originated in
the satellite galaxy whose merger with the original Milky Way disc led
to the formation of the thick disc~\citep{gwn02}; (ii) the
identification of a ring-like structure at the outer edge of the
Galactic disc~\citep{iilft03} which spans roughly $100^\circ$ in
longitude and $\pm 30^\circ$ in latitude; (iii) the detection of an
overdensity in the 2MASS all-sky M-giant catalogue which may be the
remnant of an accreted dwarf galaxy~\citep{mibild04}. The detailed
relationship between these observations is unclear. ~\cite{helmi03}
discuss a debris origin for the ring structure which has been shown by
~\cite{yanny03} to be a kinematically coherent structure. It has also
been suggested that the overdensity identified by~\cite{mibild04} was
the progenitor of the ring -- however, ~\cite{momany04} suggest that
the overdensity is instead related to the warp and flare in the
external disc. However, what is clear is that the disc of the Milky
Way has had a complicated formation history which {\it Gaia} will be
uniquely able to probe.

While these discoveries rely on observations which probe to magnitudes
fainter than the cut-off for the RVS, it is important to note that the
identification of the outer ring relied on the availability of large
area photometric surveys: earlier, small-area surveys had simply
overlooked its presence due to its relatively low surface density. The
true nature of this structure will only be confirmed once spectroscopy
is available for more of its members in order to determine its
kinematic relationship to the rest of the Milky Way disc. The
discovery of the new thick disc population was based on radial
velocities with accuracies of about $15 \kms$, a level of accuracy
which will be achieved by the RVS for all the stars it surveys. The
great strength of the RVS is that any new structures which are
identified will already have a wealth of kinematic data with which to
determine their nature. If necessary, follow-up observations from the
ground can be used to study the detailed properties of individual
structures.

\begin{figure}
\includegraphics[width=0.47\textwidth]{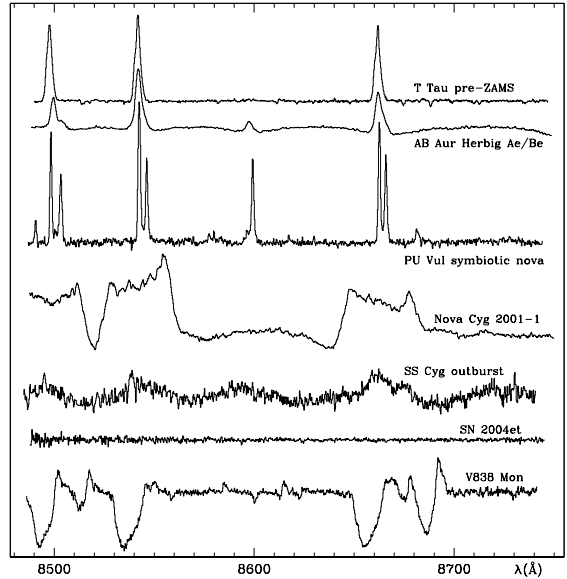}
\caption{Example spectra of common classes of outburst objects for
which {\it Gaia} will issue alerts to permit ground-based
follow-up. See text for a detailed discussion.}
\label{fig:alerts}
\end{figure}
During its five year lifetime, the {\it Gaia} satellite will observe
many transient objects, such as supernovae and microlensing
events~\citep{belokurov03}. It is intended that many such events will
be identified in real time on the ground, with alerts being issued
(within 24 hours in the case of supernovae) to enable ground-based
follow-up. RVS spectra will provide immediate classification of the
brightest photometric alerts, those of greatest interest for
follow-up. All photometric alerts with Cousin's $I_{\rm
C}$$\leq$14~mag will be bright enough for {\it Gaia} spectroscopy to
provide discriminant classification on even the single, first epoch
spectrum. Fig.~\ref{fig:alerts} shows ground-based spectra, recorded
with the Asiago 1.82m telescope operating in {\it Gaia}-like mode, of
objects most frequently appearing among those undergoing
outbursts. These spectra outline the major spectroscopic differences
to be expected between different classes of outburst which will
trigger {\it Gaia} photometric alerts.  Brightening in pre-ZAMS
objects (like T Tau and AB Aur in Fig.~\ref{fig:alerts}) can lead to
an increase of some magnitudes, accompanied by strong and wide CaII
emission (and weaker Paschen lines) from the circumstellar disc. The
width of the emission lines in novae (Nova Cyg 2001-1 in
Fig.~\ref{fig:alerts}) is very large, tracing the outward velocity of
the ejected material (of the order of 1\,000 km~sec$^{-1}$). Dwarf
novae (SS Cyg in Fig.~\ref{fig:alerts}) display a smaller width
connected with the lower Keplerian velocities in the accretion disc,
and the emission lines in symbiotic novae (PU Vul in
Fig.~\ref{fig:alerts}) are even narrower tracing the small velocities
of the wind from a late type giant suddenly ionised by the burst of
hard radiation from a white dwarf (WD) companion undergoing a
non-degenerate thermonuclear outburst. The spectra of supernovae (type
II SN 2004et in Fig.~\ref{fig:alerts}) appear as a flat continuum, the
width of the entire {\it Gaia} wavelength range corresponding to a
typical ejection velocity of 10\,000 km~sec$^{-1}$.  Finally, ejection
of optically thick material gives rise to spectacular P-Cyg profiles,
like those displayed by V838~Mon (500 km~sec$^{-1}$ terminal velocity
for the spectrum in Fig.~\ref{fig:alerts}) later to became famous for
its huge light-echo. As the Figure clearly shows, RVS spectra will be
able to distinguish between the different types of outburst thereby
facilitating decision-making about the necessity and urgency of
ground-based follow-up.

\section{Binary Stars}
\label{sec:binaries}
Another domain in which the {\it Gaia} mission will bring new and
extensive data is the detection, classification and complete
characterisation of binary systems. In this field, the RVS will be a
key contributor by providing multi-epoch measurements of the radial
velocity. It will impact on nearly all aspects of astrophysics where
binarity is a basic quantity or provides the means to derive
fundamental parameters. The statistical properties of binary stars
(binary fraction, mass ratio and period distribution) will improve our
understanding of the conditions of star formation in various locations
in the Galaxy. Masses and radii, in addition to absolute luminosities,
are fundamental inputs to stellar physics (see
\S~\ref{sec:stellar_evolution}). A knowledge of masses for a large
variety of stellar types will improve the determination of the
mass-luminosity function and directly impact on our knowledge of the
Initial Mass Function (see \S~\ref{sec:imf}).

The difficulty of obtaining unbiased samples from ground-based
observations is underlined by the significant temporal separation
of~\cite{1991A&A...248..485D} and~\cite{halbwachs03} despite being
papers II and III of the same series. In~\cite{1991A&A...248..485D},
37 spectroscopic orbits were derived and the duplicity of Solar-like
stars in the Solar neighbourhood was studied using a sample of 164
systems. \cite{halbwachs03}, which is based on two CORAVEL radial
velocity surveys and {\it Hipparcos} distances, uses an original
sample of about 600 stars.  The limitation of both papers comes from
small number statistics. For comparison, {\it Gaia} will discover
about 60 million binaries: more than $10^7$ astrometric binaries,
$10^6$ eclipsing binaries, $10^6$ spectroscopic binaries, $10^7$
resolved binaries within 250 pc. Of these, some 10\,000 will yield
stellar mass determinations accurate to within $1$ per
cent~\citep{arenouhalb02}. More specifically, the RVS will be the only
instrument on board {\it Gaia} which will be able to discover
virtually all non-eclipsing binaries with orbital periods up to a
month for stars brighter than V = 14~\citep{zm04}. These observations
will provide large and unbiased samples of binaries for a very large
range of periods and mass ratios and for stars at all evolutionary
stages (even the shortest-lived).  This will make it possible to study
period and mass ratio distributions as a function of spectral type,
population, formation site, etc.

The specific cases of eclipsing binary stars (\S~\ref{sec:eclipse}),
cataclysmic variables (\S~\ref{sec:CV}), and symbiotic stars
(\S~\ref{sec:symbiotic}) are detailed below.

\subsection{Eclipsing Binary Stars}
\label{sec:eclipse}
In the case of eclipsing systems the radial velocity information will
be combined with multi-band photometry to derive a complete orbital
solution. Note that {\it Gaia} photometry alone can be used to
discover much fainter binaries, but it will be limited to eclipsing
systems and these are uncommon for periods longer than a few days. In
addition, the RVS velocities will efficiently complement the
astrometric data for the period range $0.1 {\rm yr}$ to $5 {\rm yr}$
\citep{2003gsst.conf..351S}. The observation of eclipsing binaries by
{\it Gaia}-RVS will be of prime importance to advances in stellar
astrophysics, since stellar masses, in addition to radii and surface
temperatures, can be derived with high accuracy. The distribution of
parameters such as mass ratios, eccentricity and orbital periods can
cast light on the various processes which have been proposed for the
formation of close binary systems~\citep{2002MNRAS.332L..65B}. In
addition, there are aspects of the binary evolution models that are
not well understood.  We note that an apparent discrepancy between
observations and models for binary masses near $0.8$M$_\odot$ has been
found \citep{1997AJ....114.1195P, 2001A&A...374..980C}. In these
systems, the secondary star appears to be more evolved than the
primary in both luminosity and radius. Modelling these stars is
further complicated by the fact that many objects show evidence of
spot activity. Good samples are needed before drawing any firm
conclusions. The high resolution of the RVS spectra will allow the
derivation of the statistical properties of spectroscopic binaries
with an accuracy that cannot be achieved from ground-based surveys,
due to the required number of measurements.

{\it Gaia} will derive stellar masses to an accuracy better than $2$
per cent~\citep{marrese04,2003A&A...404..333Z,2001A&A...378..477M},
stellar radii with $1-4$ per cent accuracy, and mass ratios with $1$
per cent accuracy for more than 10$^5$ eclipsing binaries brighter
than V=15; of those at least $25$ per cent will also be double-lined
spectroscopic binaries, providing an enormous data base. In addition,
synthetic modelling of the spectra of spectroscopic binaries using
lines from both objects allows one to derive fundamental parameters
such as temperature and surface gravity which can be directly compared
with the results of the orbital solution.  Those data can be directly
compared with stellar models on the theoretical L-T$_{\rm eff}$ plane,
thereby overcoming the difficulties (for example, bolometric
corrections, colour-temperature transformations, distance, reddening)
that always hamper such a comparison on the observational V, (B-V)
plane. If the stars of a binary pair have significantly different
masses, then the requirement that the models fit the data for a single
age provides a strong constraint. An accuracy of around $1$ per cent
in mass and radius is then required to retrieve accurate information
about opacities, convection prescriptions and rotation
\citep{2002ohds.conf..187A}.  The need for additional mixing in stars
has been confirmed by several authors using binary star data.  We
note, among others, the pioneering work by \cite{1988A&A...196..128A},
the review by~\cite{2002ohds.conf..187A} , and the recent work
by~\cite{2004A&A...417.1083S} who find excellent agreement between the
Padova isochrones and the observed data on both components of V432
Aur. \cite{2001ApJ...556..230Y} present new evidence that the
overshoot efficiency might depend on the stellar mass. All these
results need to be tested on the very much larger data base that {\it
Gaia} will provide.

Knowledge of the temperatures and radii of stars accurately determines
their luminosities. Combining these with measured multi-colour
apparent magnitudes and allowing for reddening~\citep{prsa05} yields
an accurate distance to the binary~\citep{ww03}. Note that this method
of distance determination is complementary to astrometric measurements
and is limited only by the limiting magnitude of the instrument. The
method has recently been used to derive a distance to the
Pleiades~\citep{2004A&A...418L..31M} and to study the distances of
objects as far as the Large Magellanic Cloud~\citep[see][ and
references therein]{fitz03}. Results of the binary method agree well
with astrometric distances for a dozen eclipsing binaries discovered
by {\it Hipparcos}~\citep{zm04}.

\subsection{Mass Function Determination} 
\label{sec:imf}

The stellar initial mass function (IMF) is of fundamental importance
in many fields.  On the one hand, the stellar mass distribution
determines the evolution, surface brightness, chemical enrichment, and
baryonic content of galaxies. On the other hand, knowledge of the
slope of the IMF and its possible variations can cast light on the
physics of cloud fragmentation and the star formation processes
through which different mass ranges of star are assembled. 

Determining the IMF of a stellar population of mixed ages is a
cumbersome affair.  In fact, the IMF is usually obtained from the
stellar luminosity function (SLF), i.e. the number of stars in a
survey volume per magnitude interval, through knowledge of the
mass-luminosity relation (MLR) in all mass ranges and of the star
formation rate.  In the vast majority of cases, the stellar masses
cannot be directly derived and the mass has to be deduced indirectly
from the luminosity and the evolutionary state of a star.  Two basic
approaches have been tried in the literature to derive the SLF: the
first one makes use of a local volume-limited catalogue of stars with
well-measured distances. The second method takes larger samples of
stars from deep photometry. This latter is affected by various
spurious effects, such as the Malmquist bias, imprecise determination
of the completeness and unknown binary corrections
\citep{2001MNRAS.322..231K}. In particular, \cite{1995ApJ...453..358K}
has shown that the significant difference between the volume-limited
and magnitude-limited SLF for the local disc population at magnitudes
fainter than M$_{\rm V}=11.5$ is mainly due to the presence of
unresolved binary stars.  

The determination of the SLF from a volume-limited star catalogue
gives more reliable results. However, to date, complete samples of
trigonometric parallaxes are known only for stars brighter than
M$_{\rm V} \sim 9.5$ at distances d $< 20$ pc, while for the faint M
dwarfs the estimated completeness distance is 5 pc
\citep{1997A&A...325..159L}.  This means that only a limited sample of
objects is covered. A major caveat of any photometric LF is that the
determination of the distance relies on the photometric determination
from a CMD. In practice, the determination of the IMF from the SLF
requires knowledge of the chemical composition of each star, since the
absolute magnitude and colour depend on both metallicity and
age. Assuming Solar metallicity for a metal poor thick disc star would
lead to an underestimate of the absolute magnitude, an overestimate of
the distance and thus an underestimate of the number density.

The RVS will contribute to the determination of the IMF through the
accurate estimation of the binary fraction and of stellar
parameters. This, together with a parallax determination, will
precisely locate a star on the CMD and will allow a better
determination of the stellar MLR which can subsequently be used to
convert stellar luminosity functions into stellar mass functions. Our
present-day knowledge of the MLR for masses lower than $1.5$M$_{\sun}$
is based on data on visual binaries, since only a limited amount of
high quality data on double-lined eclipsing binaries and resolved
spectroscopic binaries is currently available
\citep{1997A&A...320...79M}.

By combining the proper motion, radial velocity information and
chemical abundances of single stars, information can be derived about
the dependence of the IMF on metal content. In fact, while at the
upper mass end ($>10$M$_\odot$) the IMF seems to be virtually
independent of the metal content, a dependence has been suggested at
the low mass end \citep{2001A&A...373..886R} where a flatter slope has
been found for thick disc stars. However, while on the one hand the
IMF of the metal poor stars in the Galactic spheroid does not show
this behaviour \citep{1998ApJ...503..798G}, on the other hand
\cite{2000ApJ...530..418Z} derive an IMF for the metal rich Galactic
bulge which is consistent with that of the metal poor globular
clusters.

The existence, if any, of extremely metal poor stars will cast light
on the primordial IMF. Any observed variation of the IMF between
different environments would suggest that cloud fragmentation based on
the Jeans formulation (i.e. on gravity) is not the only mechanism at
work, but rather that different processes (such as turbulence) might
be important \citep{2003IAUS..208...61K, 1999ApJ...526..279P}.

\subsection{Cataclysmic Variables}
\label{sec:CV}

Cataclysmic Variables (CVs) are a diverse class of short-period
semi-detached binaries consisting of an accreting white dwarf (WD)
primary and (typically) a low-mass, main-sequence secondary star. They
are valuable manifestations of late-stage binary evolution, which also
provide a window into the fundamental physical processes associated
with accretion and nova explosions (including, possibly, Type Ia
supernovae). The RVS spectra will be useful for identifying CVs and
will contain substantial astrophysical information on line strengths,
shapes and velocity variations. CV spectra show strong and distinctive
lines in the RVS band: broad ($\sim 3000\kms$, $8\,$nm) double-peaked
Ca lines, originating in the cooler outer edges of the disc; narrow
emission (from the heated face of the secondary star) or, in longer
period systems where the secondary is larger and brighter in this
band, photospheric absorption features instead of emission; broad
Paschen lines, one of which coincides with the Ca line at $\sim
866\,$nm. The spectra vary substantially from system to system,
depending on the orbital period, accretion rate and disc
state. Uniquely among astrophysical objects, some of these systems
show extremely strong NI lines. In magnetic CVs, the Ca triplet can be
seen strongly in emission in the one or two systems studied but very
few data are available.

For the current RVS specifications a practical limiting magnitude of
$V\sim16$ can be expected for determining CV system parameters. There
are currently 140 CVs brighter than V=16~\citep{dws01}. Extrapolating
from the initial findings of the Sloan Digital Sky
Survey~\citep{szkody02}, the incompleteness level is about $30$ per
cent on the basis of selection by colour only (which has strong
selection effects). {\it Gaia} spectroscopy will be very sensitive in
the search for signatures of accretion (line emission). This should
lead to the discovery of a large number of intrinsically faint systems
(possibly up to $1000$). All of these CVs will have excellent
parallaxes and hence luminosities. The ratios of secondary to primary
masses provide an important test of population models of interacting
binaries and can be calculated from the orbital period and velocity of
the secondary: for a significant fraction of CVs in the RVS sample,
the binary orbit and inclination will be obtained from reflection
modelling, eclipses, or directly from the {\it Gaia} astrometry.

{\it Gaia} will provide a minimally biased sample of CVs for
population and evolution studies. It should be noted that spectroscopy
from the RVS is particularly powerful: many CVs hardly vary (the
nova-likes) and many systems which would be considered detached based
on photometry are evidently accreting only once spectroscopy is
obtained.  Periods are much easier to determine with radial velocity
measurements than by photometry. The spectroscopy will also pick up
new classes of unexpected objects such as short period systems with K-
rather than M-type secondaries. There is also the likelihood of
finding longer period double-degenerate and other `graveyard' CVs, as
evolution models predict large numbers of these. In current surveys
they would be indistinguishable from WDs on the basis of their
colours. In addition, the RVS survey will determine the fraction with
Ca emission compared to absorption and the relationship of these with
the different classes of CV (magnetic/non-magnetic), luminosity and
secondary spectral type.  Ultimately, information will be gained on
the cause of the differing line strengths.

The Ca triplet lines are more clearly double-peaked than other strong
lines in the optical spectrum because they originate in the cooler
outer regions of the disc. For a large number of disc CVs, Doppler
tomography will be possible. This will produce maps of disc velocities
including the effects of tidal distortions and the infalling stream,
and the run of Ca emission within the disc. It will also identify
other aspects of disc structure, for example the spiral waves that
have been seen in some Dwarf Novae~\citep{shh97}, which have
implications for our understanding of accretion discs in general. For
those CVs for which data are available, many (perhaps most) seem to
have narrow Ca triplet emission components.  The strength of these
lines and their ratios are inputs for atmospheric heating models of
the secondary star. The Ca triplet is particularly good for both of
the above applications: Ca velocity maps are of intrinsically higher
resolution than the more commonly used H$\alpha$ maps and, because Ca
is less saturated than H$\alpha$, the associated atmospheric heating
models are simplified~\citep{marshduck96}.

\subsection {Symbiotic stars: the Supernova connection}
\label{sec:symbiotic}

Type Ia Supernovae (SN~Ia), the only supernovae to explode in
elliptical galaxies, are widely used as cosmological distance
indicators. They provide a clear indication of cosmic acceleration and
of a non-zero value for Einstein's cosmological
constant~\citep[e.g.][]{perl99}. While current search programs have
been successful in discovering new SN~Ia at high redshift (e.g. the
High-Z Supernova Search Team (HZSST) and Supernova Cosmology Project
(SCP) surveys), the true nature of the SN~Ia progenitors and their
explosion mechanisms is still a matter of investigation and
debate. With look-back times of roughly half the current age of the
Universe, one has to make sure that possible evolution of the
progenitors is not mimicking a cosmological effect. It is generally
believed that the explosion of a SN~Ia is associated with a WD that
grows in mass to the Chandrasekhar limit via accretion. Most of the
debate about the SN~Ia progenitors involves the nature of the
companion feeding mass to the WD. In the {\em single degenerate}
scenario it is a main-sequence or giant star, while in the {\em double
degenerate} scenario the companion is itself a WD and the supernova
explosion results from the merger of the two WDs in a binary system.

A possibility for the single degenerate scenario is that the donor
star is a late type giant, giving rise to a symbiotic star
(SyS)~\citep{1986syst.book.....K}. \cite{1994MmSAI..65..157M} has
listed the requirements that SyS have to possess to be viable
progenitors of SN~Ia.  These include: ($a$) their WDs must be massive
enough to be able to grow to the Chandrasekhar limit by accretion of
only a fraction of the mass reservoir transferable from their
companions; ($b$) their number in the Galaxy must be appropriate to
account for the SN~Ia rate given the evolutionary time scale of the
donor giant; ($c$) the SyS must be appropriately frequent among the
bulge/thick-disc/halo populations of the Galaxy, which correspond to
the dominant stellar populations in elliptical galaxies.

{\it Gaia} is ideally suited to address all these issues. First, {\it
Gaia} will detect most late type giants in the Galaxy and it will be
able to discover most of those harboured in SyS systems, thus
providing the fractional rates within the various Galactic
populations. SyS systems display a distinctive appearance in the RVS
wavelength range~\citep[e.g.][]{1988A&A...189...97S,
1999A&AS..140...69Z, 2002A&A...383..188M, 2003A&A...406..995M} that,
coupled with their characteristic photometric variability and spectral
energy distribution, will allow the RVS to identify them easily even
in low S/N spectra. In fact, the cool giants in Galactic SyS are
mainly of spectral types K and M, thus emitting strongly in the CaII
region, away from the veiling effect of the blue continuum emitted by
the circumstellar ionised gas. This gas also shows up in the RVS
spectral range by displaying strong Paschen, CaII and HeI emission
lines (sometimes also useful to trace the orbital motion of the
ionising WD companion). Monitoring their radial velocities over the
5~yr mission life-time (3 times longer than the typical $\sim 600$ day
orbital periods), the RVS will provide orbital periods for a
statistically significant number of objects. An indication of the mass
ratio between the cool giant and the accreting WD can be obtained by
comparing the radial velocity curve of the giant with that of the
emission lines of the highest ionisation states which presumably arise
in the regions closest to the WD. In fact, none of the strong lines of
very high ionisation levels in the spectra of SyS fall within the RVS
wavelength range. However, knowing their orbital periods and phases
from the combination of \textit{Gaia} photometry and radial
velocities, a small number of ground-based spectra at appropriate
wavelengths and phases will suffice to determine the mass ratio for
the SyS discovered by \textit{Gaia}.  These will then constrain the
mass distribution function of WDs in SyS. The \textit{Gaia}
medium-band photometric bands in the red part of the spectrum have
been designed to be sensitive to molecular bands. The intensity of
these bands is metallicity dependent, which yields an additional
constraint on the metallicity of the SyS. Finally, the location of the
SyS in the Galactic phase-space, coupled with the metallicity and
chemical analysis, will firmly address their relationship to the
various Galactic stellar populations and thereby constrain their rate
of occurrence in elliptical galaxies.

\cite{1992ApJ...397L..87M} estimated the number of SyS in the whole
Galaxy to be about $3\times 10^5$, which implies that if only $2-4$
per cent of them end their evolution in a supernova explosion it would
account for the observed SN~Ia rate. Their Galactic kinematics
\citep{1994A&A...287...87M} and infrared photometric properties
\citep{1992A&A...255..171W} argue that the majority of SyS are
associated with the bulge/thick-disc component of the Galaxy. Evidence
from energy radiated in the UV \citep{1994A&A...287...87M} and from
orbital motions suggest that a sizeable fraction of known SyS harbour
massive WDs, resembling stable H-burning conditions and only minor
mass loss. Thus, all the ingredients seem to be in place but it will
be {\it Gaia} that will provide the statistically sufficient data set
to address definitively the issue of whether or not SyS are viable
SN~Ia progenitors.

\section{Towards an understanding of stellar evolution}
\label{sec:stellar_evolution}

\subsection {Introduction: RVS impact on stellar astrophysics}
\label{sec:stell_evol_intro}
In spite of considerable recent efforts in the area of stellar
evolution, many open problems still remain.  The micro-physics of the
equation of state and opacities has been tested in the range of
parameters corresponding to the radiative zone of the Sun using
asteroseismology. However, our knowledge of macro-physics processes
such as rotation, convection and turbulence over the entire stellar
mass range remains very poor. In order to improve significantly our
knowledge of stellar interiors, it is necessary to couple the
determination of global parameters for a statistically significant
sample of stars with information from seismology
\citep{2002EAS.....2..131L,2000IAUJD..13E...4L,2000ARA&A..38...35L}.
Seismology needs a precise knowledge of the global parameters of the
stars: luminosity, effective temperature, radii or masses in the case
of binaries, surface chemical composition. In fact, seismology can be
used to derive the fundamental stellar parameters from stellar
oscillation frequencies~\citep{1997AJ....113.1457P,
1996A&A...312..463P}.  However, if global parameters are known,
stellar oscillation frequencies can be predicted and compared with the
observations, giving information about the physics of the stellar
interiors.  

{\it Gaia} astrometry, photometry and spectroscopy will build a
complete and homogeneous sample of accurate global parameters for a
large range of stellar masses. The spectra from the RVS in particular
can be used to derive the effective temperature T$_{\rm eff}$, surface
gravity $\log g$, metallicity $\rm [Fe/H]$, rotational velocity
V$_{\rm rot}$, chromospheric activity and interstellar extinction (see
Table~\ref{tab:performance} and Paper~I). When combined with parallax
determinations, the luminosity, radius and mass can be
obtained~\citep{bjones2005}. The expected accuracies of the
various parameter determinations are discussed in Paper~I (see also
Table~\ref{tab:performance}). Finally, the absolute magnitudes of
stars having distances measured with an accuracy of $1$ per cent will
be known to within $0.03$ mag. The accurate placement of stars in HR
diagrams will allow the construction of very precise stellar tracks
and give hints about the physical processes taking place in stellar
interiors.

It is well known that the location of a star in the HR diagram does
not allow a unique determination of its age, since several
combinations of [Fe/H], [$\alpha$/Fe] and age are possible. To derive
accurate ages requires very precise estimates of effective
temperatures, reddening and chemical composition.  For old main
sequence halo stars the expected RVS accuracy on the stellar
parameters (see Table~\ref{tab:performance}), combined with
information from the photometric observations, will lead to
uncertainties in individual age determinations of about $20$ per
cent. The largest contribution to this error is from the uncertainty
in [Fe/H]. However, since the age determination will be made by
comparing the location of the stars in the HR diagram with theoretical
models, further uncertainties will come from our poor knowledge of
stellar models (i.e. mixing, diffusion, nuclear rates, etc.). The
stellar parameter determinations from \textit{Gaia} will be therefore
first be used as input to improve our knowledge of stellar models.

In the following sections, we underline some of the areas where the
RVS will particularly contribute to our understanding of stellar
physics.

\subsection {Star formation and pre-main-sequence evolution}

The process of star formation in different environments is far from
being understood. In particular, it is difficult to reconcile the
prominent influence of the local environment (turbulence, compression,
initial trigger) on small scales with the universality of the Schmidt
and Kennicut law on Galactic scales~\citep[see][ for a recent
review]{2002ApJ...577..206E} which suggests that Galactic-scale
gravity is involved in the first stages of star formation. The {\it Gaia}
study of Galactic star forming regions will shed light on the details
and modality of star formation both in space (does the star formation
take place in giant molecular clouds or in small cloudlets?), and in
time (from the age distribution of stars) deriving absolute
luminosities, effective temperatures, multiplicities, kinematics and,
where possible, masses of the star forming complexes.

Many aspects need to be clarified concerning the pre-main-sequence
(PMS) evolution of stars.  In the domain of small masses ($\leq
1.5$M$_\odot$) the main uncertainties are related to the treatment of
over-adiabatic convection, the stars being fully convective at the
beginning and developing a small radiative core during the H-burning
phase.  The PMS tracks run almost parallel to the main sequence,
starting from the deuterium burning phase which represents the
starting point of the evolution in the visible. Unless binaries are
present, mass estimates rely on the assumptions of the convective
model \citep{1999BaltA...8..253D}. The prototypical PMS star of this
mass range is T Tauri. During those initial stages, parameters such as
stellar rotation, angular momentum evolution and magnetic field
strength play very important roles which need to be quantified by
observed data.

Massive stars are believed to be formed via continuous mass accretion
during the PMS phase~\citep{1998stne.conf..101P}. The star follows the
birth-line accreting mass and, when the accretion phase stops, the
object leaves the birth-line and moves towards the main sequence. This
model is in agreement with several observational constraints. However,
not everything is well explained. The location of the birth-line is
strongly dependent on the accretion rate which is essentially unknown.
The accretion rate greatly influences all the internal properties of
the stars as well as the PMS lifetime \citep{2002Ap&SS.281...75M}.
Additionally, the accretion rate derived for low-mass stars fails to
describe massive objects, since the formation time would be too long
compared to the main-sequence lifetime \citep{2000A&A...359.1025N}:
stars would leave the main sequence before being fully formed.

An additional source of uncertainty for massive PMS stars is related
to their high rotation velocities and to the treatment of convection
(see the following Sections for a detailed discussion). Accurate
distance determinations for very young associations and clusters,
together with spectroscopic determination of the effective
temperatures, rotational velocities and infrared spectra of PMS stars
will lead to more accurate PMS tracks.

The angular momentum budget of stars and the way in which stars lose
it has not yet been fully understood.  PMS stars are known to be fast
rotators.  In young clusters there seems to be a relation between the
ages and rotational velocities of the stars, in the sense that stars
appear to be losing angular momentum \citep{2000AJ....119.1303T,
2003PASP..115..505S} during their lifetime.  The transfer of angular
momentum through the star and the role of the proto-stellar disc
predicted by the models need to be compared with observations.

A large number of (proto)-stars in the PMS phase are likely to be
detected by {\it Gaia}, with a significant sample of all ages ranging
from the earliest T-Tauri phase (few Myr) to the onset of central
H-burning (several tens of Myr).  {\it Gaia} will observe about 120
young open clusters like Praesepe within 1 kpc of the Sun with an
accuracy comparable to that reached by {\it Hipparcos} in the Hyades
where low mass PMS stars are visible. In the RVS wavelength range, the
spectra of T-Tauri stars show all three CaII lines in very strong
emission. This will allow for the identification of T-Tauri stars even
at very low S/N ratios.

\subsection {Rotation}
\label{sec:rotation}
Despite the fact that many observations have demonstrated that
rotation is a necessary ingredient in models of massive stars
\citep{2000ARA&A..38..143M,2001ApJ...563..334S}, the effect of stellar
rotation on the evolution of stars has only recently been included in
stellar models. Among the critical observations, we recall the fact
that rapidly rotating massive O stars have peculiar He abundances. B-
and A-type stars in the Magellanic Clouds are found to have large
relative excesses of He/H with respect to the prediction of current
models without rotation. The most likely explanation of these peculiar
abundances is rotation-induced mixing. Rotation can influence the
location of a star in the HR diagram, its luminosity and its lifetime.
The evolution of massive stars is the result of an interplay between
rapid rotation and mass loss. In fact, massive objects are rapidly
rotating when they form and as a result of the reduced internal
pressure they are sub-luminous.  During the main-sequence lifetime the
stars decelerate owing to angular momentum loss through their winds
and become more luminous more rapidly than non-rotating stars
\citep{2000ARA&A..38..143M}.  Rotation increases the H-burning time
for stars more massive than 9~M$_\odot$ by about $25-30$ per cent,
while the effect on the He-burning lifetime is smaller~\citep[about
$10$ per cent:][]{2000ARA&A..38..143M}.  The effect on age
determinations is far from being negligible, amounting to about $25$
per cent.  In that sense, rotation can mimic the effects of core
overshoot.  To date, only 20\,000 stars have measured $v \sin i$
\citep{2000AcA....50..509G}.  A detailed analysis has revealed that
while early type stars, from O-type to early F, are fast rotators
($50-400\kms$), late type objects (from late F to M) possess lower
rotational velocities ($v \sin i \leq 50\kms$;
\citeauthor{2001BaltA..10..613M}, \citeyear{2001BaltA..10..613M},
\citeauthor{2001ApJ...563..334S}, \citeyear{2001ApJ...563..334S}). PMS
stars are known to be fast rotators reaching 30 per cent of the
breakup velocity -- as they age, mass loss causes them to lose angular
momentum. In fact, in young star clusters, a relation has been found
between the age of the cluster and the distribution of rotational
velocities of the
stars~\citep{2000AJ....119.1303T,2002ApJ...576..950T}. {\it Gaia} will
study $120$ young clusters closer than $1$ kpc, allowing the
determination of the rotational velocity. The interaction of the
convective envelope of low mass stars with their differential rotation
sustains a magnetic field which, by trapping the ionised stellar
winds, induces a loss of angular momentum from the stars.

The RVS will greatly improve our knowledge by directly measuring
rotation via line broadening and spot transit for a large sample of
stars: the combination of accurate luminosity, effective temperature
and rotational velocity determinations provided by the RVS spectra
will cast light on their evolution.  \cite{2003gsst.conf..285G} and
\cite{gomboc05} discuss the accuracy of the determination of $v \sin
i$ from the cross-correlation of the RVS spectra of single stars.
Precisions of $v \sin i \sim$ 5 km/s should be obtained at the end of
the mission for late type stars down to V=15. For B5 main-sequence
stars, we expect an accuracy of $v \sin i \sim$ 10-20 km/s up to
V$\sim$ 10-11.  The periodic transit of spots on the stellar surface
can also trace the rotation. The RVS spectra can trace several spots
at once as they cross the projected surface. The associated emission
lines will split according to the velocity, allowing accurate rotation
measurements even with only a few tens of
spectra~\citep{2003gsst.conf.....M}.
 
\subsection {Mixing processes}
One of the main points of uncertainty in stellar evolution theory
concerns the extent of the convectively unstable regions and
associated mixing.  Thermal convection arises in stars when radiation
is not sufficient to carry the heat flux originating from the deep
interior. The convective instability occurs wherever the local
temperature gradient is steeper than the adiabatic gradient, a
condition called the Schwarzschild criterion.
 
However, an extra mixing beyond the classical convective regions
(overshoot) seems to be at work~\citep[see][ for a
review]{1999sstt.conf....9C}.  The theory of non-local convection has
made significant progress over the past decade
\citep{1990A&A...232...31X, 1999ApJ...518L.119C,
1998ApJ...493..834C,1992ApJ...389..724C,
1996MNRAS.279..305G,1998A&A...334..953V, 1999ApJ...518L.119C,
2000ApJ...534L.113C, 2002ApJ...570..825B,
2002AAS...200.0715Y}. However, the lack of satisfactory results means
that the majority of stellar models are still calculated using the
local approach of mixing length theory (from the pioneering work by
\cite{1975A&A....40..303M} and \cite{bressan81} to the recent work of
\cite{2002A&A...391..195G}), where the mean free path of the
convective elements is proportional to the scale height of the
pressure H$_{\rm p}$ through a free parameter derived from comparisons
with the data.  Overshoot can greatly influence the evolution of
massive and intermediate-mass stars, changing the lifetimes in the H-
and He-burning phases, as well as the luminosity of the stars. The
effect on the age determination of a stellar population is far from
being negligible and is about $25$ per cent for an A star of age $2$
Gyr.  Element diffusion can also change the evolution of a star, which
can introduce errors as large as 100 Myr in the age of a
$1.7$M$_\odot$ star. As far as low mass stars are concerned, in the
mass range where the core switches from the radiative to the
convective regime the determination of the overshoot parameter leads
to unsatisfactory results \citep{2001AJ....122.1602W}. This
corresponds to the age range of the oldest open clusters (6 Gyr),
introducing large uncertainties on the age determination itself. By
combining the expected {\it Gaia} determinations of the global
parameters of stars with the results of asteroseismology measurements
(see \S~\ref{sec:stell_evol_intro}), it has been estimated that when L
and T$_{\rm eff}$ of stars are known with a precision of about $2$ per
cent, then the overshoot parameter can be derived with a precision of
0.03 times the pressure scale height H$_{\rm p}$~\citep
{2000ARA&A..38...35L}.

A discussion of the lines of observational evidence for the presence
of overshoot is given by \cite{1998IAUS..189..323C}. The analysis of
the HR diagram of {\it Hipparcos} field stars shows that the region of
the so-called Hertzsprung gap is very sensitive to overshoot on the
main sequence for stars with masses of 1.6 M$_\odot$
\citep{1998A&A...334..901S}.  Using the data to infer the history of
star formation in the disc, \cite{2001AJ....121.1013B} find that the
compatibility between the number of main-sequence and
post-main-sequence stars requires some extra mixing during the core
H-burning phase in stars in the range 1.5-2 M$_\odot$. Information
about the size of the convective core can be derived from m$_{\rm He}$
the maximum mass undergoing a core He-flash.  m$_{\rm He}$ can be
derived from the luminosity function of the red clump stars: the mass
distribution inside the clump presents a peak at m$_{\rm He}$ where
the He-burning time in the core has a maximum
\citep{2001MNRAS.323..109G}. The ratio $^{12}$C/$^{13}$C in AGB stars
can be derived from RVS spectroscopy and gives hints about the third
dredge-up, which in turn depends on the mass loss rate and on the
treatment of convection~\citep{2000A&A...360..617M}.

Classical Cepheids provide quantitative constraints on the efficiency
of mixing phenomena (rotational mixing, overshoot, semi-convection)
among intermediate-mass stars \citep{2000ARA&A..38..143M,cassisi2004}.
The discrepancy between evolutionary masses and pulsation masses for
Classical Cepheids dates back to the 1970s. Recent investigations
suggest that this discrepancy is at the level of $10$ per cent for
Galactic Cepheids~\citep{2001ApJ...563..319B} but of the order of
$15-20$ per cent for Magellanic
Cepheids~\citep{2001A&A...373..164B}. We still lack clear physical
arguments to explain whether this discrepancy is either a real feature
or a consequence of errors in the physical assumptions adopted to
construct evolutionary and pulsation
models~\citep{2002ApJ...565L..83B}. Suggestions have been advanced in
the literature that the Cepheid mass discrepancy can be alleviated by
including overshoot~\citep{1992ApJ...387..320C}.

The careful calibration of the HR diagram as a function of age and
metallicity by {\it Gaia} will give very important hints about stellar
interiors and, when considered in conjunction with the results of
seismology observations, will improve our capability to construct
non-local convective models.

\subsection {Mass loss}

Mass loss from stars is important for the evolution of the
interstellar medium whose composition depends on the nature of the
supplied material. Stars are known to lose mass at various rates
during different stages of their evolution.  The RVS observations will
be well adapted for the study of stellar winds and mass loss, since
CaII and H lines display characteristic P-Cyg profiles in the RVS
spectral range even for modest mass loss rates
\citep{2003gsst.conf.....M}.

In particular, mass loss is a dominant effect in the upper main
sequence \citep{1986ARA&A..24..329C}. The mass loss rates currently
used are based on observations \citep{1988A&AS...72..259D,
1996fstg.conf..162L}.  However, the relation between rotation and mass
loss is not yet properly quantified.  While \cite{1985ApJ...299..255V}
finds a significant increase in mass loss with rotation for OB stars,
only a modest increase is derived by \cite{1988A&A...203..355N}. The
same uncertainty is present on the theoretical side. Whether or not
mass loss influences the angular momentum seems to depend on whether
it is proceeding via equatorial or polar winds~\citep[see][ for a
review]{2000A&A...361..159M}.

In the domain of low mass stars, it is not yet clear whether the
pulsation instability might drive the efficiency of mass-loss close to
the RGB tip.  There is mounting empirical evidence that a substantial
fraction of bright RGB stars are
variables~\citep{1996ApJ...464L.157E}.  This result was strongly
supported by detailed analysis of time-series data collected by
microlensing experiments~\citep{2003MNRAS.343L..79K,wood2004} as well
as by Near-Infrared surveys~\citep{2002MNRAS.337L..31I}. This is a new
and very interesting result because it has been suggested that these
objects are excited stochastically by
convection~\citep{2001ApJ...562L.141C}. The unique opportunity
provided by {\it Gaia} to measure distances and to collect homogeneous
multi-band photometric and spectroscopic data for a very large sample
of halo and disc RGB stars is crucial to provide a comprehensive
analysis of these objects. In particular, the RVS will supply firm
constraints on the role played by chemical composition and binary
companions in the pulsational de-stabilisation of K-type stars.

\subsection {Variable stars}
Variable stars are typically used as primary distance indicators or as
stellar tracers in the Galaxy as well as in external galaxies.
However, they also play a fundamental role in stellar astrophysics,
because they can be easily identified and their radial velocity curves
are key observables to estimate fundamental parameters such as stellar
masses, radii, and effective temperatures. Therefore, it is not
surprising that variable stars are also adopted to constrain basic
physics problems such as an upper limit to the neutrino magnetic
moment ~\citep{1990ApJ...365..559R,1993ApJ...402..574C} or the
existence of dark matter
particles~\citep{1990ApJ...354..568D,2000PhR...333..593R}. Moreover,
they also provide the unique opportunity to compare predictions based
on hydrodynamical pulsation models, evolutionary models, and stellar
atmosphere models.  {\it Gaia} will play a paramount role in the field
of variable stars and this topic has already been discussed in several
papers~\citep[e.g.][ and references
therein]{tammann02,bono03a,bono03b}. Alongside trigonometric parallax
determinations, {\it Gaia} will provide photometric and spectroscopic
time series data for a large number of variable stars throughout the
Galactic spheroid. This will create a unique opportunity to address,
on a quantitative basis, several long-standing problems. In the
following, we briefly mention some of the key issues.

\subsubsection{Intermediate-mass Variable Stars}   
\label{sec:inter_mass_var}
Recent theoretical~\citep{bono99b,bono99a} and
empirical~\citep{tammann03,kervella04} results indicate that the
Cepheid Period-Luminosity (PL) relation is not universal. In fact,
both the zero-point and the slope of the optical PL relation seem to
depend on the metal content. Moreover, detailed investigations of
Galactic and Magellanic fundamental Cepheids~\citep{Sandage04,kanbur04}
support theoretical predictions~\citep{bono99b,bono99a} concerning the
nonlinearity of optical PL relations when moving from short to
long-periods. This has not been conclusively established, however,
since different theoretical predictions based on linear Cepheid
models~\citep{baraffe01} and empirical approaches based on the
Infrared-Flux method~\citep{luck03,storm04} suggest a mild dependence
on the metal content. More recently, theoretical predictions based on
nonlinear, convective models and multi-band observations suggest that
the PL relation of first-overtone Galactic and Magellanic Cepheids
marginally depends on metal abundance~\citep{2002ApJ...565L..83B}.

\begin{figure}
\includegraphics[width=0.5\textwidth]{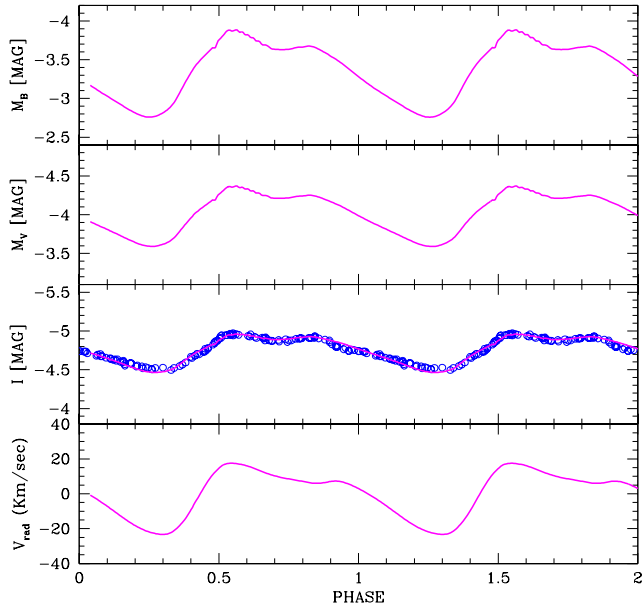}
\caption{Predicted radial velocity and light curves in different
photometric bands for a Large Magellanic Cloud Bump Cepheid. A
comparison with empirical I-band data is also shown (third panel),
taken from OGLE data.}
\label{fig:cepheid_fig}
\end{figure}

{\it Gaia} will supply a new impetus to studies of the Cepheid
distance scale because it will obtain an almost complete census of
Galactic Cepheids. Classical Cepheids in the period range from 6 to 16
days present a well-defined bump in both their luminosity and radial
velocity curves. This feature, for periods shorter than about $9$
days, is located along the decreasing branch, while for longer periods
it is located along the rising branch. Fig.~\ref{fig:cepheid_fig}
presents predicted radial velocity and light curves in different
photometric bands for one such Bump Cepheid in the Large Magellanic
Cloud. As the Figure illustrates, the velocity changes will be
detectable by the RVS at a statistically significant level down to a
magnitude limit of about $V=14$, where the velocity errors are below
$6\kms$ for single epoch observations (Paper~I). Accurate chemical
compositions (intrinsic error smaller than 0.2 dex) will also be
obtained down to a limiting magnitude of
$V\approx12-13$~\citep{mun03,2003gsst.conf..291T}.

We note that the amplitude of radial velocity variations exhibited by
Classical Cepheids ranges from $15$ to $90\kms$ as one moves from
short period to long period objects~\citep[see
e.g.][]{1999AcA....49..201U,2003ApJ...599..522P}. By assuming an
absolute magnitude of M$_{\rm V}\approx-2$ for short-period Cepheids,
the RVS will thus provide accurate measurements of chemical
composition, reddening, and pulsation properties for all Cepheids
located in the outer disc out to distances of about $6-10$ kpc
(depending on reddening) and over the entire Galactic disc for the
brightest ones. 

The current observational data set of metallicity measurements for
Galactic and Magellanic Cepheids is quite limited.  Accurate
spectroscopic measurements of Galactic Cepheids have been recently
provided by~\citeauthor{FryCarney97} (\citeyear{FryCarney97}, 23
objects) and by~\citeauthor{andrievsky04} (\citeyear{andrievsky04},
131 objects). But only for three dozen of these Cepheids are accurate
absolute distance determinations also
available~\citep{storm04}. Accurate spectroscopic measurements of
Magellanic Cepheids are available for 40
objects~\citep{luck98,romaniello2005}, but accurate distance
determinations are available only for a few of
them~\citep{gieren00}. Moreover, optical PL relations based on
nonlinear convective models predict that, on moving from metal-poor
(Z=0.004, the metal abundance typical of Small Magellanic Cloud
Cepheids) to metal-rich (Z=0.02, the metal abundance typical of
Galactic Cepheids), there is a difference in magnitude that at $\log P
=1$ is 0.4 mag in the $V-$band and 0.3 mag in the
$I-$band~\citep{fiorentino02}. Empirical estimates of this effect
still present a broad range of values. Using the RVS data, it will be
possible to perform a robust calibration of the optical PL and
Period-Luminosity-Colour (PLC) relations~\citep[for a more
quantitative discussion of both random and systematic errors
see][]{bono2005}. To illustrate this point, let us assume that the
metallicity dependence is roughly equal to $M_{\rm V} / \delta \log Z
\approx \mid0.4\mid $. A series of random extractions based on the
OGLE catalogue~\citep{udalski01} indicates that a sample of 1500
Cepheids with metal abundance ranging from Z=0.002 to Z=0.02 and for
which {\em i)} geometric absolute distances with an accuracy better
than $2$ per cent ($\sim 0.04$ in distance modulus); {\em ii)} metal
abundances with an accuracy better than 0.2 dex; {\em iii)} reddening
estimates with an accuracy better than 0.02 mag are available will
allow us not only to constrain the metallicity dependence in the
optical bands with an accuracy better than 0.01 magnitude but also to
constrain the fine structure of both the PL and PLC relations over the
entire period range ($0.3 \le \log P \le 1.8$).  At the same time, the
comparison between theory and observations will supply firm
constraints on the accuracy and plausibility of the physical
assumptions adopted in the construction of hydrodynamical models of
variable stars.

Classical Cepheids located in the outer disc are important for the
distance scale and to estimate disc metallicity gradients and the
shape and scale of Galactic rotation \citep{pqbm97,mcs98}. Note that
we know the chemical composition for only two Cepheids located at
distances larger than $16$ kpc~\citep{luck03} and the radial
velocities for just a handful of Cepheids in the Galactic
anti-centre. The recent discovery of an extended HI spiral arm located
in the fourth quadrant of the Milky Way at a radial distance of
$18-24$ kpc highlights the potential value of a large-volume sample of
Cepheid tracers in the Galactic disc~\citep{mcclure04}.
 
\subsubsection{Low-mass Variable Stars}   

RR Lyrae stars are the most popular low-mass distance indicators. When
compared with Classical Cepheids they are at least 2-3 mag fainter but
they are ubiquitous in the Galactic spheroid and in dwarf galaxies.
The optical luminosity and metallicity of RR-Lyrae stars are related
through an M$_{\rm V}-$[Fe/H] relation.  Different aspects of this
relation have been addressed in countless empirical and theoretical
investigations ~\citep[see][ and references
therein]{1999glcl.conf.....M,2003sced.conf...85B,2003sced.conf..105C,catelan2003}.
The calibration of this relation is still affected by uncertainties in
distance estimates, reddening corrections and evolution off the
zero-age horizontal branch (ZAHB).  However, recent theoretical and
empirical evidence indicates that the K-band
Period-Luminosity-Metallicity relation is only marginally affected by
these deceptive
uncertainties~\citep{2003MNRAS.344.1097B}. \cite{2002AJ....123..473B}
recently provided an accurate trigonometric parallax for RR Lyr itself
based on observations obtained with the Fine Guide Sensor on board HST
(although they had only 5 reference stars). This notwithstanding, a
robust calibration of both the zero-point and slope of the M$_{\rm V}$
vs $\rm [Fe/H]$ relation still awaits better statistics. If we assume
an absolute magnitude of M$_{\rm V}\approx0.6$ at $\rm
[Fe/H]=-1.5$~\citep{2003sced.conf..105C}, we find that the RVS will
provide accurate measurements of chemical composition, reddening, and
pulsation properties for the entire sample of metal-poor ($\rm
[Fe/H]=-1.65\pm0.03$:~\citeauthor{1994ApJS...93..271S},
~\citeyear{1994ApJS...93..271S}), halo RR Lyrae out to distances of
about $2-3$ kpc.

A further interesting aspect for which the RVS will be valuable is the
detection and measurement of RR Lyrae stars in binary systems.  We
still lack detailed measurements of the dynamical masses of RR Lyrae,
since at present there are only a few
Galactic~\citep{1999AJ....118.2442W} and Large Magellanic
Cloud~\citep{2003AcA....53...93S} candidates.

{\it Gaia} will also create the unique opportunity to calibrate the PL
and the PLC relation for the known Type II Cepheids, namely stars of
type BL Herculis, W Virginis and RV Tauri. At present, the empirical
relations are hampered by small number statistics because only a
handful of such stars have been identified in Globular
Clusters~\citep{2003AJ....126.1381P} and in the
LMC~\citep{1998AJ....115.1921A}. These objects are very good distance
indicators because they are at least $1$ mag brighter than RR Lyrae
stars.  In addition, theory and observations suggest that their
properties depend only marginally on metallicity~\citep{bcs97}.

\section{Conclusions}
The Radial Velocity Spectrometer (RVS) on board the {\it Gaia}
satellite will provide spectra in the 848 nm to 874 nm wavelength
range with a resolving power R = 11\,500 for stars brighter than
magnitude $V=17$ over approximately 90 percent of the sky. The
unbiased nature of this data set, combined with the multi-epoch
observations of each object, render the RVS data set unique among
spectral surveys of the Galaxy. In this paper, we have highlighted the
many areas in which the RVS spectra constitute an essential complement
to the astrometric and photometric data which {\it Gaia} will
collect. For studies of Galactic structure and evolution, the unbiased
sample of radial velocities obtained from the RVS spectra will provide
the sixth phase-space coordinate which is vital for the construction
of well-constrained Galactic models. The multi-epoch nature of the
observations will allow the identification of large numbers of binary
stars in the {\it Gaia} data set while the availability of spectra
taken simultaneously with the photometric observations will be
invaluable for the study of detailed stellar properties, including
stellar variability.

The {\it Gaia} mission, and the RVS in particular, will generate a
data set of unprecedented size and precision which will revolutionise
our understanding of all aspects of Galactic astronomy and stellar
astrophysics. Harnessing the full power of such a data set presents
many challenges, both practical and theoretical. For example, as we
discussed in \S~\ref{sec:movinggroups}, the {\it Hipparcos} results
for the local velocity distribution emphasise the difficulty of
disentangling the global properties of the disc distribution function
from small scale structure such as moving groups. Given this
difficulty, it will not necessarily be fruitful, or even possible, to
model the {\it Gaia} data set via a simple decomposition into known
populations such as disc, bulge and halo due to the ambiguity in the
assignment of individual stars to specific populations. Instead, a
global approach will be needed, which makes simultaneous use of
spatial and kinematic information to model the entire data set as a
coherent unit~\citep[e.g.][]{binney05}. During the coming decade,
prior to the satellite launch, much work needs to be done in order to
determine the iterative Galactic modelling schemes which will make
optimal use of the {\it Gaia} data set and will lead us to a coherent
interpretation of the vast quantity of information it will
contain. Similar developments will be required for the study of
binaries and stellar astrophysics in the post-{\it Gaia} era.

\subsection*{Acknowledgements}
We thank the RVS consortium for their on-going work on the simulation,
development and testing of the RVS instrument. MIW thanks PPARC for
financial support. DK, FT, FA, CT and SM acknowledge financial support
from CNES. It is a pleasure to thank P.B. Stetson for providing us
with the photometric data for the three standard stellar fields
adopted in Fig.~\ref{fig:disc_fields} as well as C.E. Corsi and
G. Iannicola for the sending us the photometric data adopted in
Fig.~\ref{fig:disc_fields} in advance of publication. GB acknowledges
partial support by MIUR COFIN~2003 and by INAF2003. GB also thanks
M. Romaniello for several interesting discussions concerning Cepheid
metal abundances. We also thank the anonymous referee for helpful
comments.

\end{document}